\documentclass[prb,twocolumn,showpacs,amsmath,amssymb]{revtex4}

\usepackage{graphicx}
\usepackage{dcolumn}
\usepackage{bm}

\begin{document}
\title
{Revised Calculation of $\bm \kappa_2(T)$ for $s$-wave Type-II Superconductors}

\author{Takafumi Kita}
\affiliation{Division of Physics, Hokkaido University, Sapporo 060-0810,
Japan}

\date{\today}

\begin{abstract}
This paper presents revised calculations for the Maki parameters $\kappa_{1}$ and $\kappa_{2}$ and the pair potential $\Delta({\bf r})$ of $s$-wave type-II superconductors near the upper critical field $H_{c2}$ with arbitrary impurity concentration. It is found that Eilenberger's well-known results on $\kappa_{2}$ [Phys.\ Rev.\ {\bf 153},\ 584 (1967)] are not correct quantitatively, which are modified appropriately. Calculations are also performed for a two-dimensional system with an isotropic Fermi surface. The results in the clean limit differ substantially from those in three dimensions with an spherical Fermi surface. This fact indicates the necessity of considering detailed Fermi-surface structures for a quantitative understanding of the parameters. The coefficient of $\Delta({\bf r})\!\propto\!(H_{c2}\!-\!B)^{1/2}$, which is basic to any theoretical evaluation of the thermodynamic and transport properties near $H_{c2}$, is obtained accurately.
\end{abstract}
\maketitle

\section{Introduction}
\label{sec:intro}

Following the preceding studies,\cite{Gor'kov59-2,Maki64,deGennes65,
Tewordt65,HW66,WHH66,Eilenberger66,MT65,NT66,CCdG66,Maki66}
Eilenberger\cite{Eilenberger67}
performed an extensive calculation of the parameters $\kappa_{1}(T)$ and
$\kappa_{2}(T)$ introduced by Maki\cite{Maki64} 
to distinguish temperature dependences of the upper critical field $H_{c2}$
and the initial slope of the magnetization $\partial M/\partial H$, respectively.
Based on the $s$-wave pairing with a spherical Fermi surface
and taking both $s$- and $p$-wave impurity scatterings into account,
he clarified a basic feature that $\kappa_{2}\!\geq\! \kappa_{1}\!\geq\! \kappa_{\rm GL}$,
where $\kappa_{\rm GL}$ is the Ginzburg-Landau parameter near $T_{c}$.\cite{GL50,FH69}
He also found a large dependence of the parameters on the $p$-wave scattering strength.
This study is undoubtedly one of the basic works in the field 
and has been referred to frequently
in analyzing experimental results on the quantities.
It will be shown, however, that his results on $\kappa_{2}$
are not correct quantitatively
due to a couple of inappropriate approximations adopted.

This fact also tells us that we are still far from a quantitative description
of type-II superconductors.
The parameter $\kappa_{2}$ is such a basic quantity that it is relevant 
to all thermodynamic and transport properties near $H_{c2}$.
Indeed, changes of those quantities through $H_{c2}$ are proportional to 
the spatial average $\langle|\Delta({\bf r})|^{2}\rangle$ of
the pair potential $\Delta({\bf r})$, and $\langle|\Delta({\bf r})|^{2}\rangle$ 
is directly connected with $\kappa_{2}$, as seen below.
Thus, an absence of a reliable theory on $\kappa_{2}$ also implies no quantitative
theories for all the other quantities near $H_{c2}$.
The exact limiting behaviors would be useful not only for their own sake,
but also for getting an insight into the behaviors over $0\!\leq\! B \!\leq\! H_{c2}$.
In addition, they would serve as a guide for any detailed numerical studies
for $0\!\leq\! B \!\leq\! H_{c2}$.

With these observations, I here perform revised calculations for the Maki parameters
and the pair potential near $H_{c2}$.
Besides correcting Eilenberger's results on $\kappa_{2}$ for the spherical Fermi surface, 
I also perform two-dimensional calculations of the quantities for an isotropic 
(i.e.\ cylindrical or circle) Fermi surface.
Thereby clarified will be a rather large dependence of
$\kappa_{1}(T)$ and $\kappa_{2}(T)$
on detailed Fermi-surface structures.
Indeed, even the empirical inequality $\kappa_{2}\!\geq\! \kappa_{1}\!\geq\! \kappa_{\rm GL}$
will be shown violated in some cases in two dimensions,
even without the spin paramagnetism.\cite{Maki66}

The starting point adopted for these purposes
is the quasiclassical Eilenberger equations.\cite{Eilenberger68}
As emphasized by Eilenberger\cite{Eilenberger68} and also by Serene and Rainer,\cite{SR83}
the quasiclassical equations have an advantage over Gor'kov equations\cite{Gor'kov59}
that they are easier to solve due to the absence of an irrelevant energy variable.
They have a rigorous microscopic foundation and hence form a firm basis
for any quantitative description of superconducting/superfluid Fermi liquids.\cite{SR83}
Thus, it seems somewhat surprising that few calculations on $\kappa_{2}$
have been performed based on the Eilenberger equations.\cite{OB75}

This paper is organized as follows. 
Section \ref{sec:formulation} provides the
formulation for the $s$-wave pairing with an isotropic Fermi surface
and $s$-wave impurity scattering,
deferring $p$-wave impurity scattering to Appendix \ref{sec:p-wave}.
The main results are given in Secs.\ \ref{subsec:solutions} and \ref{subsec:KN'N},
and the differences from Eilenberger's calculation are
explained in Sec.\ \ref{subsec:diffEilen}.
Section \ref{sec:results} presents numerical results.
Section \ref{sec:summary} summarizes the paper,
with possible extensions
to include realistic Fermi surfaces from first-principles calculations
and/or anisotropic pairings.
Appendix \ref{sec:H-B} derives an analytic expression for the magnetization.

\section{\label{sec:formulation}formulation}

\subsection{Eilenberger equations}

I consider the $s$-wave pairing with an isotropic Fermi surface and $s$-wave 
impurity scattering in an external magnetic field ${\bf H}\parallel \!{\bf z}$.
The vector potential in the bulk can be written 
as\cite{Eilenberger64,Lasher,Marcus,Brandt,DGR90,Kita98}
\begin{equation}
{\bf A}({\bf r})\!=\! Bx\hat{\bf y}\!+\tilde{\bf A}({\bf r})\, ,
\label{A}
\end{equation}
where $B$ is the average flux density produced jointly by the external current 
outside the sample and the supercurrent inside it, and
$\tilde{\bf A}$ expresses the spatially varying part of the magnetic field 
satisfying $\int{\bm \nabla}\!\times\!\tilde{\bf A}\,d{\bf r}\!=\!{\bf 0}$.
I adopt the units where the energy, the length, and the magnetic field
are measured by the zero-temperature energy gap $\Delta(0)$ at $H\!=\!0$,
the coherence length
$\xi_{0}\!\equiv\!\hbar v_{\rm F}/\Delta(0)$ with $v_{\rm F}$ the Fermi velocity, 
and $B_{0}\!\equiv\!\phi_{0}/2\pi\xi_{0}^{2}$ with 
$\phi_{0}\!\equiv\! hc/2e$ the flux quantum,
respectively. I also put $\hbar\!=\!k_{\rm B}\!=\! 1$ and 
use the gauge where ${\bf \nabla}\cdot\tilde{\bf A}\!=\!0$.
The Eilenberger equations\cite{Eilenberger68} now read
\begin{subequations}
\label{Eilens}
\begin{eqnarray}
\left[\varepsilon_{n}+\frac{\langle g\rangle}{2\tau}
+\frac{\hat{\bf v}}{2}\!\cdot\!\left({\bm \nabla}-i{\bf A}\right)\right] f=
\left(\Delta+\frac{\langle f\rangle}{2\tau}\right)g \, ,
\label{Eilen}
\end{eqnarray}
\begin{eqnarray}
\Delta({\bf r}) \ln \!\frac{T_{c}}{T}= -2\pi T \sum_{n=0}^{\infty}
\left[\langle 
f(\varepsilon_{n},{\bf k}_{\rm F},{\bf r})\rangle -\frac{\Delta({\bf r})}{\varepsilon_{n}}\right] \, ,
\label{pair}
\end{eqnarray}
\begin{eqnarray}
-{\nabla}^{2}\tilde{\bf A}({\bf r})\!=-\frac{i}{\kappa_{0}^{2}}
2\pi T \sum_{n=0}^{\infty}\langle \hat{\bf v}\,g(\varepsilon_{n},{\bf k}_{\rm F},{\bf r})\rangle \, .
\label{Maxwell}
\end{eqnarray}
\end{subequations}
Here $\varepsilon_{n}\!\equiv\!(2n+1)\pi T$ is the Matsubara frequency, 
$\tau$ is the relaxation time in the second-Born approximation,
$\langle \cdots \rangle$ denotes
the Fermi-surface average satisfying $\langle 1 \rangle\!=\! 1$, 
$\Delta({\bf r})$ is the pair potential, and
the unit vector $\hat{\bf v}\!=\!\hat{\bf k}$ 
specifies a point on the isotropic Fermi surface.
The quasiclassical Green's functions $f$ and $g$ are connected by
$g\!=\!(1\!-\! ff^{\dagger})^{1/2}$
with $f^{\dagger}(\varepsilon_{n},{\bf k}_{\rm F},{\bf r})\!\equiv
\!f^{*}(\varepsilon_{n},-{\bf k}_{\rm F},{\bf r})$, and
the dimensionless parameter $\kappa_{0}$ is defined by
\begin{equation}
\kappa_{0}\!\equiv\! {\phi_{0}}/{2\pi\xi_{0}^{2} H_{c}(0)}\, ,
\label{kappa0Def}
\end{equation}
where $H_{c}(0)\!\equiv\!\sqrt{4\pi N(0)}\Delta(0)$ is the thermodynamic critical
field at $T\!=\!0$ with $N(0)$ the density of states per spin and per unit volume.
Equations (\ref{Eilen})-(\ref{Maxwell}) are to be solved self-consistently for a fixed $B$.
Finally, the missing connection between $H$ and $B$ is obtained by applying
the Doria-Gubernatis-Rainer scaling \cite{DGR89} to Eilenberger's
free-energy functional. \cite{Eilenberger68}
The details are given in Appendix A. 
The final result is given by
\begin{eqnarray}
&&\hspace{-3mm}H=B+\frac{1}{BV}\int\! d{\bf r}\,({\bm \nabla}\!\times\!\tilde{\bf A})^{2}
\nonumber \\
&&\hspace{-3mm}+\frac{\pi T}{2BV\kappa_{0}^{2}}\!\sum_{n=0}^{\infty}\int \! d{\bf r} \left<\!
\frac{f^{\dagger}\hat{\bf v}\!\cdot\!({\bm \nabla}\!-\!i{\bf A})f
\!-\!f\hat{\bf v}\!\cdot\!({\bm \nabla}\!+\!i{\bf A})f^{\dagger}}{1+g}\!\right>,
\nonumber \\
\label{H-B}
\end{eqnarray}
where $V$ is the volume of the system.

\subsection{Expansion near $H_{c2}$}

Near $H_{c2}$, the coupled equations (\ref{Eilens}) and (\ref{H-B}) 
are expanded in terms of $\Delta({\bf r})$ as follows.
First, let us rewrite\cite{Kita98}
\begin{equation}
\frac{\hat{\bf v}}{2}\!\cdot\!\left({\bm \nabla}-i{\bf A}\right)
=\frac{\sqrt{B}\sin\theta}{2\sqrt{2}}\left[{\rm e}^{-i\varphi}(a\!+\!\tilde{A})
-{\rm e}^{i\varphi}(a^{\dagger}\!+\!\tilde{A}^{*})\right],
\label{nabla-a}
\end{equation}
where $(\theta,\varphi)$ are the polar angles of $\hat{\bf v}$, and
the quantities $a$, $a^{\dagger}$, and $\tilde{A}$ are defined by
\begin{subequations}
\begin{equation}
\left\{
\begin{array}{l}
\vspace{2mm}
\displaystyle
a\equiv \frac{1}{\sqrt{2B}}\left(\frac{\partial}{\partial x}
+i\frac{\partial}{\partial y}+Bx\right)
\\
\displaystyle
a^{\dagger}\equiv \frac{1}{\sqrt{2B}}\left(-\frac{\partial}{\partial x}
+i\frac{\partial}{\partial y}+Bx\right)
\end{array}
\right. ,
\label{a}
\end{equation}
\begin{equation}
\tilde{A}\!\equiv\!-i  \frac{\tilde{A}_{x}+i\tilde{A}_{y}}{\sqrt{2B}} \, ,
\label{Atilde}
\end{equation}
\end{subequations}
with $[a,a^{\dagger}]\!=\!1$.
The operators $(a,a^{\dagger})$ are the same as 
$(a_{-},a_{+})$ introduced by Helfand and Werthamer,\cite{HW66} and $(F_{-},F_{+})$
by Eilenberger.\cite{Eilenberger67}

I then expand
$f$, $g$, and $\tilde{A}$ up to the third order in $\Delta({\bf r})$ as
\begin{equation}
\left\{
\begin{array}{l}
\vspace{2mm}
f=f^{(1)}+f^{(3)}
\\
\vspace{2mm}
g= 1- \frac{1}{2}f^{(1)\dagger}f^{(1)}
\\ 
\tilde{A}\!=\!\tilde{A}^{(2)}
\end{array}
\right. .
\label{fgAExp}
\end{equation}
Substituting Eqs.\ (\ref{nabla-a}) and (\ref{fgAExp}) into Eq.\ (\ref{Eilen})
and collecting terms of the same orders, we obtain
\begin{subequations}
\label{f(1)-(3)}
\begin{equation}
\bigl[\,\tilde{\varepsilon}_{n}
+\beta\,({\rm e}^{-i\varphi}a
-{\rm e}^{i\varphi}a^{\dagger})\,\bigr] f^{(1)}=\Delta+\frac{\langle f^{(1)}\rangle}{2\tau}\, ,
\label{f(1)0}
\end{equation}
\begin{eqnarray}
&&\hspace{-9mm}\bigl[\,\tilde{\varepsilon}_{n}
+\beta\,({\rm e}^{-i\varphi}a
-{\rm e}^{i\varphi}a^{\dagger})\,\bigr] f^{(3)}=\frac{\langle f^{(3)}\rangle}{2\tau}
\nonumber \\
&&\hspace{-5mm}
-\frac{f^{(1)\dagger}f^{(1)}}{2}\left(\!\Delta+\frac{\langle f^{(1)}\rangle}{2\tau}\!\right)
+\frac{\langle f^{(1)\dagger}f^{(1)}\rangle}{4\tau} f^{(1)}
\nonumber \\
&&\hspace{-5mm}
-\beta\,({\rm e}^{-i\varphi}\tilde{A}^{(2)}
-{\rm e}^{i\varphi}\tilde{A}^{(2)*})f^{(1)}
\, ,
\label{f(3)0}
\end{eqnarray}
\end{subequations}
with
\begin{equation}
\tilde{\varepsilon}_{n}\equiv \varepsilon_{n}+\frac{1}{2\tau}, \hspace{5mm} 
\beta\equiv \frac{\sqrt{B}\sin\theta}{2\sqrt{2}} .
\label{beta}
\end{equation}
Also, Eq.\ (\ref{pair}) is transformed into
\begin{equation}
\Delta({\bf r}) \ln \!\frac{T_{c}}{T}= -2\pi T \sum_{n=0}^{\infty}
\left[\langle 
f^{(1)}\rangle +\langle 
f^{(3)}\rangle-\frac{\Delta({\bf r})}{\varepsilon_{n}}\right] \, .
\label{D(3)}
\end{equation}
The Maxwell equation (\ref{Maxwell}) is given in the leading order by
\begin{equation}
-\nabla^{2}\tilde{A}^{(2)}=\frac{\pi T}{\sqrt{2B}\kappa_{0}^{2}}\sum_{n}
\langle f^{(1)\dagger}f^{(1)}{\rm e}^{i\varphi}\sin\theta \rangle \, ,
\label{A(2)}
\end{equation}
whereas Eq.\ (\ref{H-B}) becomes
\begin{eqnarray}
H-B=\frac{\pi T}{2\sqrt{2B}\kappa_{0}^{2}V}\sum_{n}
\int \! d{\bf r}\,\langle f^{(1)\dagger}({\rm e}^{-i\varphi}a
\nonumber \\
-{\rm e}^{i\varphi}a^{\dagger}) f^{(1)}\sin\theta\rangle\, .
\label{H-B(2)}
\end{eqnarray}

\subsection{\label{subsec:2d}Two-dimensional calculations}

To investigate the dependence of the Maki parameters on the Fermi-surface structure,
calculations will also be performed for a two-dimensional system
with an isotropic Fermi surface placed
in the $xy$ plane perpendicular to ${\bf B}$.
The analytic expressions for this case can be obtained 
from those of the three dimensions by 
simply putting $\sin\theta\!\rightarrow\! 1$ and omitting the integrations over $\theta$.

\subsection{Transformation into algebraic equations}

Equations (\ref{f(1)-(3)}) and (\ref{D(3)})-(\ref{H-B(2)}) are solved 
with the Landau-level expansion (LLX) method\cite{Kita98}
by expanding $\Delta$, $f^{(\nu)}$ $(\nu\!=\! 1,3)$, and $\tilde{A}^{(2)}$ 
in terms of periodic basis functions of the flux-line lattice
as
\begin{subequations}
\label{Expand}
\begin{equation}
\Delta({\bf r})
=\sqrt{V}
\sum_{N=0}^{\infty}\Delta_{N}\,\psi_{N{\bf q}}({\bf r}) \, ,
\label{DeltaExpand}
\end{equation}
\begin{equation}
f^{(\nu)}(\varepsilon_{n},{\bf k}_{\rm F},{\bf r})
=\sqrt{V}\sum_{m=-\infty}^{\infty}
\sum_{N=0}^{\infty}f_{mN}^{(\nu)}(\varepsilon_{n},\theta)\,
{\rm e}^{i m\varphi}\,\psi_{N{\bf q}}({\bf r}) 
\, ,
\label{fExpand}
\end{equation}
\begin{equation}
\tilde{A}^{(2)}({\bf r}) = \sum_{{\bf K}\neq {\bf 0}} \tilde{A}_{\bf K}^{(2)}\, 
{\rm e}^{i{\bf K}\cdot{\bf r}} \, .
\label{AExpand}
\end{equation}
\end{subequations}
Here $N$ denotes the Landau level, 
${\bf q}$ is an arbitrary chosen magnetic Bloch vector 
characterizing the broken translational symmetry of the flux-line lattice
and specifying the core locations,
and ${\bf K}$ is a reciprocal lattice vector of the magnetic Brillouin zone.
See Ref.\ \onlinecite{Kita98} for the explicit expressions of 
the basis functions $\psi_{N{\bf q}}({\bf r})$;
they are essentially equivalent to Eilenberger's 
$\psi_N({\bf r}|{\bf r}_{0})$ \cite{Eilenberger67} and reduces for $N\!=\! 0$
to Abrikosov's solution for the Ginzburg-Landau equations
near $H_{c2}$.\cite{Abrikosov57}
It only suffices to know the properties:
\begin{subequations}
\begin{equation}
\langle\psi_{N{\bf q}}|\psi_{N'{\bf q}}\rangle=\delta_{NN'} \, ,
\end{equation}
\begin{equation}
\left\{
\begin{array}{l}
\vspace{2mm}
\displaystyle a\, \psi_{N{\bf q}}\,=\sqrt{N}\,\psi_{N-1{\bf q}}
\\
\displaystyle a^{\dagger} \psi_{N{\bf q}}=\sqrt{N\!+\! 1}\,\psi_{N+1{\bf q}}
\end{array}
\right. .
\end{equation}
\end{subequations}
On the other hand, the expansion in ${\bf K}$ in Eq.\ (\ref{AExpand})
was introduced by Brandt\cite{Brandt} for solving the Ginzburg-Landau equations
over $0\!\leq\! B \!\leq\! H_{c2}$.
This expansion enables us to integrate the Maxwell equation appropriately
so that $\int{\bm \nabla}\!\times\!\tilde{\bf A}\,d{\bf r}\!=\!{\bf 0}$
is satisfied automatically.

A simplification results in Eq.\ (\ref{Expand})
for the $s$-wave pairing
with an isotropic Fermi surface near $H_{c2}$.
Indeed, $\Delta({\bf r})$ 
can be described excellently with only the lowest Landau level as\cite{Kita98}
\begin{equation}
\Delta({\bf r})
=\sqrt{V}\Delta_{0}\,\psi_{0{\bf q}}({\bf r}) \, .
\label{DeltaExpand2}
\end{equation}
Equation (\ref{DeltaExpand2}) has a wide range of applicability over $B\! \agt\!0.1H_{c2}$
both near $T_{c}$ and in the dirty limit.
However, the region in the clean limit shrinks
as $T\!\rightarrow\! 0$ to disappear eventually.
It should also be noted that 
higher Landau levels of $N\!=\!{\rm even}$ become relevant for anisotropic pairings and/or
anisotropic Fermi surfaces at low temperatures.

Substituting Eqs.\ (\ref{fExpand}),  (\ref{AExpand}), and (\ref{DeltaExpand2})
into Eq.\ (\ref{f(1)-(3)}) and using the orthogonality of ${\rm e}^{im\varphi}$
and $\psi_{N{\bf q}}$, we realize that 
$f_{mN}^{(\nu)}$ can be written as
\begin{equation}
f_{mN}^{(\nu)}\!=\!\delta_{mN}\Delta_{0}^{\nu} \tilde{f}_{N}^{(\nu)}\, .
\end{equation}
Equations (\ref{f(1)-(3)}) and (\ref{D(3)})-(\ref{H-B(2)}) 
are thereby transformed into algebraic equations for 
$\tilde{f}^{(1)}$, $\tilde{f}^{(3)}$, $\Delta_{0}$, $\tilde{A}_{\bf K}^{(2)}$,
and $H\!-\! B$ as
\begin{subequations}
\begin{equation}
\sum_{N'} \tilde{\cal M}_{NN'}\tilde{f}_{N'}^{(1)} = \delta_{N0}\left(1
+\frac{\langle \tilde{f}_{0}^{(1)}\rangle}{2\tau}\right),
\label{Eilen2-1}
\end{equation}
\begin{equation}
\sum_{N'} \tilde{\cal M}_{NN'}\tilde{f}_{N'}^{(3)} 
= \delta_{N0}\frac{\langle \tilde{f}_{0}^{(3)}\rangle}{2\tau}+J_{N}^{(3)}+J_{N}^{(A)}\, ,
\label{Eilen2-3}
\end{equation}
\begin{equation}
\ln \frac{T_{c}}{T}=-2\pi T \sum_{n=0}^{\infty}
\left(\langle 
\tilde{f}_{0}^{(1)}\rangle +\langle 
\tilde{f}_{0}^{(3)}\rangle \Delta_{0}^{2}-\frac{1}{\varepsilon_{n}}\right) ,
\label{pair2}
\end{equation}
\begin{equation}
\tilde{A}^{(2)}_{\bf K}=-\frac{2\pi T\Delta_{0}^{2}}{\kappa_{0}^{2}K^{2}}
\sum_{n=0}^{\infty}\sum_{N}\frac{J_{N}^{(2)}(\varepsilon_{n})I_{N+1N}({\bf K})}{\sqrt{N\!+\!1}} \, ,
\label{Maxwell2}
\end{equation}
\begin{equation}
H-B=\frac{2\pi T\Delta_{0}^{2}}{\kappa_{0}^{2}}
\sum_{n=0}^{\infty}\sum_{N}J_{N}^{(2)}(\varepsilon_{n}) \, .
\label{H-B2}
\end{equation}
\end{subequations}
Here the matrix $\tilde{\cal M}$ is defined by
\begin{equation}
\tilde{\cal M}_{NN'}\equiv\delta_{NN'}\tilde{\varepsilon}_{n}
+\delta_{N,N'-1}\beta\sqrt{N\!+\!1}
-\delta_{N,N'+1}\beta\sqrt{N}\, ,
\label{Matrix}
\end{equation}
and $J_{N}$'s are given by
\begin{subequations}
\label{J}
\begin{equation}
J_{N}^{(2)}\equiv\frac{(-1)^{N}}{B}\sqrt{N\!+\!1}\,
\langle\tilde{f}^{(1)}_{N+1}\tilde{f}^{(1)}_{N}\beta\rangle
\, ,
\label{J(2)}
\end{equation}
\begin{eqnarray}
&&\hspace{-3mm}J_{N}^{(3)}
\equiv-\frac{1}{2}\!\sum_{N'}(-1)^{N'} I_{NN'N+N'0}^{(4)}\,
\tilde{f}_{N'}^{(1)}\tilde{f}_{N+N'}^{(1)}\!\left(\! 1
\!+\!\frac{\langle \tilde{f}_{0}^{(1)}\rangle}{2\tau}\! \right)
\nonumber \\
&&\hspace{8mm}+\frac{1}{4\tau}\sum_{N'}(-1)^{N'}I_{NN'N'N}^{(4)}\,
\langle \tilde{f}_{N'}^{(1)}\tilde{f}_{N'}^{(1)}\rangle\tilde{f}_{N}^{(1)}
 \, ,
\label{J(3)}
\end{eqnarray}
\begin{eqnarray}
J_{N}^{(A)}\equiv-\frac{\beta}{\Delta_{0}^{2}}
\sum_{{\bf K}\neq {\bf 0}} \bigl[I_{N+1N}^{*}({\bf K})\tilde{A}_{\bf k}^{(2)}\tilde{f}_{N+1}^{(1)}
\nonumber \\
-I_{NN-1}({\bf K})\tilde{A}_{\bf k}^{(2)*}\tilde{f}_{N-1}^{(1)}\bigr]\, ,
\label{J(A)}
\end{eqnarray}
\end{subequations}
with\cite{Kita98}
\begin{subequations}
\begin{eqnarray}
I_{N_{1}N_{2}N_{3}N_{4}}^{(4)}\equiv V\int\psi_{N_{1}{\bf q}}^{*}\psi_{N_{2}{\bf q}}^{*}
\psi_{N_{3}{\bf q}}\psi_{N_{4}{\bf q}} 
d{\bf r}\, ,
\label{I4}
\end{eqnarray}
\begin{eqnarray}
I_{N_{1}N_{2}}({\bf K})\equiv
\int\psi_{N_{1}{\bf q}}^{*}\psi_{N_{2}{\bf q}}{\rm e}^{-i{\bf K}\cdot{\bf r}}
d{\bf r}\, .
\label{I2K}
\end{eqnarray}
\end{subequations}

Equation (\ref{Eilen2-1}) tells us that $\tilde{f}_{N}^{(1)}$ is real;
this fact has been used in writing down Eqs.\ (\ref{Eilen2-3})-(\ref{H-B2}).
As for $\tilde{f}_{N}^{(3)}$,
numerical calculations show that $I^{(4)}$'s appearing in Eq.\ (\ref{J(3)}) are
all real for the relevant hexagonal lattice, with 
\begin{equation}
\beta_{\rm A}\equiv I^{(4)}_{0000} = 1.16\, .
\label{betaA}
\end{equation}
Also, $I_{N+1N}({\bf K})$ can be transformed with partial integrations as
\begin{eqnarray}
&&I_{N+1N}({\bf K})=\frac{1}{\sqrt{N+1}}\int (a^\dagger \psi_{N{\bf q}})^{*}
\psi_{N{\bf q}}{\rm e}^{-i{\bf K}\cdot{\bf r}}
d{\bf r}
\nonumber \\
&& = \frac{\sqrt{N}}{\sqrt{N+1}}I_{NN-1}({\bf K})
+\frac{K_{y}\!-\!iK_{x}}{\sqrt{2B(N+1)}}I_{NN}({\bf K})
\nonumber \\
&& = \frac{K_{y}\!-\!iK_{x}}{\sqrt{2B(N\!+\!1)}}
\sum_{N_{1}=0}^{N}I_{N_{1}N_{1}}({\bf K}) \, .
\label{I2K2}
\end{eqnarray}
Since $I_{NN}({\bf K})$ is real,
$J^{(A)}$ is also real from Eq.\ (\ref{Maxwell2}), 
and so is $\tilde{f}_{N}^{(3)}$.

It is desirable for a later purpose to express
$J^{(A)}$ 
in terms of $I^{(4)}$ rather than $I_{N+1N}({\bf K})$.
This can be performed
by first substituting Eq.\ (\ref{Maxwell2}) into Eq.\ (\ref{J(A)}),
and then using Eq.\ (\ref{I2K2}) and the identity
$\sum_{{\bf K}\neq{\bf 0}}{\rm e}^{i{\bf K}\cdot({\bf r}-{\bf r}')}\!
=\!V\delta({\bf r}\!-\!{\bf r}')\!-\!1$.
The result is given by
\begin{eqnarray}
J_{N}^{(A)}=\frac{\pi T \beta}{B\kappa_{0}^{2}}
\sum_{n'=0}^{\infty}\sum_{N'}J_{N'}^{(2)}
\bigl[\sqrt{N\!+\! 1}\,\tilde{f}^{(1)}_{N+1}({\cal I}_{NN'}-1)
\nonumber \\
-\sqrt{N}\tilde{f}^{(1)}_{N-1}({\cal I}_{N-1N'}-1)\bigr],
\label{J(A)2}
\end{eqnarray}
where $\tilde{f}^{(1)}_{N\pm 1}\!\equiv\!\tilde{f}^{(1)}_{N\pm 1}(\varepsilon_{n})$,
$J_{N'}^{(2)}\!\equiv\!J_{N'}^{(2)}(\varepsilon_{n'})$
is given by Eq.\ (\ref{J(2)}),
and ${\cal I}_{NN'}$ is an average of $I^{(4)}$ defined by
\begin{equation}
{\cal I}_{NN'}\equiv\frac{1}{(N+1)(N'+1)}\sum_{N_{1}=0}^{N}
\sum_{N_{1}'=0}^{N'}I^{(4)}_{N_{1}N_{1}'N_{1}'N_{1}} \, .
\label{CalI}
\end{equation}

\subsection{\label{subsec:solutions}Solutions}

We are now ready to solve Eqs.\ (\ref{Eilen2-1}) and (\ref{Eilen2-3}).
To this end, let us define 
\begin{eqnarray}
\tilde{K}^{N'}_{N}\equiv (\tilde{\cal M}^{-1})_{NN'}=(-1)^{N+N'}\tilde{K}^{N}_{N'} \, ,
\label{KN'N}
\end{eqnarray}
where the second equality originates from 
$\tilde{\cal M}_{NN'}\!=\!(-1)^{N+N'}\tilde{\cal M}_{N'N}$.
Then Eq.\ (\ref{Eilen2-1}) is transformed into
\begin{equation}
\tilde{f}_{N}^{(1)}={\tilde{K}^{0}_{N}}
\left(1+\frac{\langle \tilde{f}_{0}^{(1)}\rangle}{2\tau}\right) \, .
\label{f(1)a}
\end{equation}
Solving Eq.\ (\ref{f(1)a}) self-consistently
for $\langle\tilde{f}_{0}^{(1)}\rangle$ and substituting 
the result into Eq.\ (\ref{f(1)a}), we obtain
\begin{equation}
\tilde{f}_{N}^{(1)}=\frac{{\tilde{K}^{0}_{N}}}{{1-\langle \tilde{K}^{0}_{0}\rangle/2\tau}} \, .
\label{f(1)}
\end{equation}
The denominator in Eq.\ (\ref{f(1)}) corresponds to the so-called ``vertex correction.''
Equation (\ref{Eilen2-3}) for $\tilde{f}_{N}^{(3)}$ may be handled similarly.
Using the symmetry $\tilde{K}^{N}_{0}\!=\!(-1)^{N}\tilde{K}^{0}_{N}$,
we thereby arrive at the expression for the relevant quantity 
$\langle \tilde{f}_{0}^{(3)}\rangle$ in Eq.\ (\ref{pair2}) as
\begin{eqnarray}
&&\langle \tilde{f}_{0}^{(3)}\rangle
=\sum_{N}(-1)^{N}\langle\tilde{f}_{N}^{(1)}(J_{N}^{(3)}\!+\!J_{N}^{(A)})\,\rangle
\, .
\label{f(3)}
\end{eqnarray}
The quantity
$\sum_{N}(-1)^{N}\langle\tilde{f}_{N}^{(1)}J_{N}^{(A)}\rangle$ in
Eq.\ (\ref{f(3)})
may be transformed further by using Eq.\ (\ref{J(A)2}) and (\ref{J(2)}) as
\begin{eqnarray}
&&\hspace{-5mm}
\sum_{N}(-1)^{N}\langle\tilde{f}_{N}^{(1)}(\varepsilon_{n})J_{N}^{(A)}(\varepsilon_{n})\rangle
\nonumber \\
&&\hspace{-8mm}
=\frac{1}{\kappa_{0}^{2}}\sum_{N}J_{N}^{(2)}(\varepsilon_{n})\, 2\pi T
\!\!\sum_{n'=0}^{\infty}\sum_{N'}J_{N'}^{(2)}(\varepsilon_{n'})({\cal I}_{NN'}\!-\!1 )
\, .
\label{sum-J(A)}
\end{eqnarray}

We next consider the self-consistency equation (\ref{pair2}) 
for the pair potential.
Here, $\tilde{f}_{N}^{(1)}(B)$ is expanded in terms of 
the distance $H_{c2}\!-\!B$ from $H_{c2}$ as
\begin{equation}
\tilde{f}_{N}^{(1)}(B)=\tilde{f}_{N}^{(1)}(H_{c2})-
\tilde{f}_{N}^{(1)\prime}(H_{c2})(H_{c2}\!-\! H\! +\! H\!-\! B) \, ,
\label{f(1)expand}
\end{equation}
whereas the higher-order term $\tilde{f}_{N}^{(3)}$ is evaluated at $H_{c2}$.
To find an explicit expression for
$\tilde{f}_{N}^{(1)\prime}$ in Eq.\ (\ref{f(1)expand}),
let us differentiate Eq.\ (\ref{Eilen2-1}) with respect to $B$:
\begin{eqnarray}
&&\hspace{-10mm}\sum_{N'} \tilde{\cal M}_{NN'}\tilde{f}_{N'}^{(1)\prime} 
= \delta_{N0}\frac{{\langle \tilde{f}_{0}^{(1)\prime}\rangle}}{2\tau}
\nonumber \\
&&\hspace{10mm}-\frac{\beta}{2B}(\sqrt{N\!+\! 1}\tilde{f}_{N+1}^{(1)}-\sqrt{N}\tilde{f}_{N-1}^{(1)})\, .
\label{Eilen2-1D}
\end{eqnarray}
This equation can be solved in the same way 
as Eq.\ (\ref{Eilen2-1}) to yield
\begin{equation}
\langle\tilde{f}_{0}^{(1)\prime}\rangle = -\sum_{N}J_{N}^{(2)} \, ,
\label{f(1)'2}
\end{equation}
where $J_{N}^{(2)}$ is defined by Eq.\ (\ref{J(2)}).
Let us substitute Eqs.\ (\ref{f(3)})-(\ref{f(1)expand}) and (\ref{f(1)'2}) into
Eq.\ (\ref{pair2}), replace $H\!-\!B$ by the right-hand side of Eq.\
(\ref{H-B2}), and regard
$H_{c2}\!-\!H$ as second order.
Collecting first-order terms,
we obtain the equation to fix the upper critical field $H_{c2}$ as
\begin{equation}
\ln \frac{T_{c}}{T}=-2\pi T \sum_{n=0}^{\infty}
\left[\langle 
\tilde{f}_{0}^{(1)}(\varepsilon_{n})\rangle-\frac{1}{\varepsilon_{n}}\right] .
\label{pair3-1}
\end{equation}
The third-order terms determine the pair potential $\Delta_{0}$ and 
the magnetization $H\!-\!B$ as a function of $H_{c2}\!-\!H$ as
\begin{subequations}
\label{MainResults}
\begin{equation}
\Delta_{0}^{2}= 
\frac{H_{c2}\!-\!H}{\kappa_{0}^{2}S_{4}/S_{2}^{2}-S_{A}/S_{2}^{2}} \, 
\frac{\kappa_{0}^{2}}{S_{2}}\, ,
\label{pair3-3}
\end{equation}
\begin{equation}
H-B= \frac{H_{c2}\!-\!H}{\kappa_{0}^{2}S_{4}/S_{2}^{2}-S_{A}/S_{2}^{2}}
\equiv\frac{H_{c2}\!-\!H}{(2\kappa_{2}^{2}-1)\beta_{\rm A}}\, ,
\label{H-B3}
\end{equation}
\end{subequations}
where $S_{2}$, $S_{4}$, and $S_{A}$ are defined by
\begin{subequations}
\label{S}
\begin{equation}
S_{2}\equiv 2\pi T\sum_{n=0}^{\infty}\sum_{N=0}^{\infty}
J_{N}^{(2)}(\varepsilon_{n})\, ,
\label{S2}
\end{equation}
\begin{equation}
S_{4}\equiv -2\pi T\sum_{n=0}^{\infty}\sum_{N=0}^{\infty}(-1)^{N}
\langle\tilde{f}_{N}^{(1)}(\varepsilon_{n})
J_{N}^{(3)}(\varepsilon_{n})\rangle \, ,
\label{S4}
\end{equation}
\begin{equation}
S_{A}\equiv (2\pi T)^{2}\sum_{n=0}^{\infty}\sum_{n'=0}^{\infty}\sum_{N=0}^{\infty}
\sum_{N'=0}^{\infty}
J_{N}^{(2)}(\varepsilon_{n})J_{N'}^{(2)}(\varepsilon_{n'}){\cal I}_{NN'}
\, ,
\label{SA}
\end{equation}
\end{subequations}
with $\tilde{f}_{N}^{(1)}$, $J_{N}$, 
and ${\cal I}_{NN'}$
given by Eqs.\ (\ref{f(1)}), (\ref{J}),
and (\ref{CalI}), respectively.
All the quantities in Eq.\ (\ref{S}) are to be evaluated at $H_{c2}$.
The latter equality in Eq.\ (\ref{H-B3}) defines the Maki parameter $\kappa_{2}$
with $\beta_{\rm A}\!\equiv\!I^{(4)}_{0000}\!=\! 1.16$.

Equation (\ref{MainResults}) forms the main result of the paper,
which is not only exact but also convenient for numerical calculations.
An extension to include $p$-wave impurity scattering is carried out in Appendix B,
where it is shown that Eq.\ (\ref{MainResults}) is still valid with the replacements of
$\tilde{f}^{(1)}_{N}$ and $J^{(3)}_{N}$ by Eqs.\ (\ref{f(1)p2}) and (\ref{J(3)p2}),
respectively.

Sometimes it is physically more meaningful to express $\Delta_{0}$ 
as a function of $B$ instead of $H$,
because $B$ is the real average field inside the bulk
directly relevant to the spatial profile of the pair potential.
It is obtained, without the replacement of $H\!-\!B$ mentioned
above Eq.\ (\ref{pair3-1}), as
\begin{equation}
\Delta_{0}^{2}= 
\frac{H_{c2}\!-\!B}{\kappa_{0}^{2}S_{4}/S_{2}^{2}-S_{A}/S_{2}^{2}+1} \, 
\frac{\kappa_{0}^{2}}{S_{2}}\, .
\label{pair3-3b}
\end{equation}

\subsection{\label{subsec:KN'N}Calculation of $\tilde{K}^{N'}_{N}$}

The key quantity in Eq.\ (\ref{MainResults}) 
is $\tilde{K}^{N'}_{N}$ defined by Eq.\ (\ref{KN'N}),
as may be seen from Eqs.\ (\ref{f(1)}),
(\ref{J}), and (\ref{S}).
An efficient algorithm to calculate them is obtained as follows.

Let us define ${\cal D}_{N}$ ($\bar{{\cal D}}_{N}$)
for $N\!=\!0,1,2,\cdots$ as the determinant of the submatrix 
obtained by removing (retaining) the first $N$ rows and columns 
of the tridiagonal matrix $\tilde{\cal M}$ of Eq.\ (\ref{Matrix}), namely,
\begin{subequations}
\begin{equation}
{\cal D}_{N}\equiv \det\!
\left[
\begin{array}{cccc}
\tilde{\varepsilon}_{n} & \sqrt{N\!+\!1}\,\beta &0 &\cdots\\
-\sqrt{N\!+\!1}\,\beta & \tilde{\varepsilon}_{n} & \sqrt{N\!+\!2}\,\beta & \cdots\\
0 & -\sqrt{N\!+\!2}\,\beta & \tilde{\varepsilon}_{n} & \cdots\\
\cdots & \cdots & \cdots & \cdots\\
\end{array}
\right],
\label{CalD}
\end{equation}
\begin{equation}
\bar{\cal D}_{N}\equiv \det\!
\left[
\begin{array}{ccccc}
\tilde{\varepsilon}_{n} & \sqrt{1}\,\beta &\cdots &\cdots&\cdots\\
-\sqrt{1}\,\beta & \tilde{\varepsilon}_{n} & \cdots & \cdots&\cdots\\
\cdots & \cdots & \cdots & \cdots&\cdots\\
\cdots & \cdots &\cdots& \tilde{\varepsilon}_{n} & \sqrt{N\!-\!1}\,\beta \\
\cdots & \cdots  &\cdots& -\sqrt{N\!-\!1}\,\beta & \tilde{\varepsilon}_{n} \\
\end{array}
\right].
\label{barCalD}
\end{equation}
\end{subequations}
They satisfy
\begin{subequations}
\begin{equation}
{\cal D}_{N-1}=\tilde{\varepsilon}_{n}{\cal D}_{N}+ N\beta^{2}{\cal D}_{N+1}\, ,
\end{equation}
\begin{equation}
{\bar {\cal D}}_{N+1}=\tilde{\varepsilon}_{n}{\bar {\cal D}}_{N}
+N\beta^{2}{\bar {\cal {\cal D}}}_{N-1} \, ,
\end{equation}
\end{subequations}
as shown by expanding Eqs.\ (\ref{CalD}) and (\ref{barCalD})
with respect to the first and last row, respectively. \cite{Aitken56}
Then $\tilde{K}^{N'}_{N}$ for $N'\!\leq\! N$ is obtained by also using standard 
techniques to solve linear equations \cite{Aitken56} as 
\begin{equation}
K^{N'}_{N}\!=\!\beta^{N-N'}\sqrt{\frac{N!}{N'!}}\,
\frac{{\cal D}_{N+1}{\bar {\cal D}}_{N'}}{{\cal D}_{0}}\, .
\end{equation}

This algorithm can put into a more convenient form in terms of
\begin{subequations}
\begin{equation}
{\cal R}_{N}\equiv \tilde{\varepsilon}_{n}{\cal D}_{N+1}/{\cal D}_{N}\, , 
\end{equation}
\begin{equation}
\bar{\cal R}_{N}\equiv \tilde{\varepsilon}_{n}\bar{\cal D}_{N-1}/\bar{\cal D}_{N}\, .
\end{equation}
\end{subequations}
They satisfy
\begin{subequations}
\label{R_N}
\begin{equation}
{\cal R}_{N-1}=\frac{1}{1+Nb^{2}{\cal R}_{N}} \, ,
\label{R_Na}
\end{equation}
\begin{equation}
\bar{\cal R}_{N+1}=\frac{1}{1+Nb^{2}\bar{\cal R}_{N}} \, ,
\label{R_Nb}
\end{equation}
\end{subequations}
with $\bar{\cal R}_{1}\!=\! 1$ and
\begin{equation}
b\equiv\beta/\tilde{\varepsilon}_{n} \, .
\label{b}
\end{equation}
Then $\tilde{K}^{N}_{N}$ and $\tilde{K}^{N'}_{N}$ for $N'\!<\! N$ are obtained by
\begin{subequations}
\label{KN'N2}
\begin{equation}
\tilde{K}^{0}_{0}={\cal R}_{0}/\tilde{\varepsilon}_{n}\, , 
\label{K00}
\end{equation}
\begin{equation}
\tilde{K}^{N}_{N}=({\cal R}_{N}/\bar{\cal R}_{N})\tilde{K}^{N-1}_{N-1}\, , 
\label{KNN}
\end{equation}
\begin{equation}
\tilde{K}^{N'}_{N}=\sqrt{N}\,b\,{\cal R}_{N}\tilde{K}^{N'}_{N-1}\, . 
\label{KN'N3}
\end{equation}
\end{subequations}
Numerical calculations of ${\cal R}_{N}$ may be carried out
by starting from ${\cal R}_{N_{\rm{cut}}}=1$
for an appropriately chosen large $N_{\rm{cut}}$
and using Eq.\ (\ref{R_Na}) to decrease $N$.
One can check the convergence by increasing $N_{\rm cut}$.
It turns out that
$N_{\rm cut}\!=\! 1$ is sufficient both near $T_{c}$
and in the dirty limit, thereby reproducing
the analytic results by Gor'kov \cite{Gor'kov59} and Caroli, Cyrot, and de Gennes,\cite{CCdG66}
respectively.
In contrast, $N_{\rm cut}\!\agt\! 1000$
is required in the clean limit at low temperatures.

Noting that Eq.\ (\ref{pair3-1}) with Eq.\ (\ref{f(1)}) should be equivalent 
to the equation for $H_{c2}$
obtained by Helfand and Werthamer,\cite{HW66}
we get an alternative expression for $\tilde{K}^{0}_{0}$ as
\begin{eqnarray}
\tilde{K}^{0}_{0}(\tilde{\varepsilon}_{n},\beta) = 
\sqrt{\frac{2}{\pi}}\int_{0}^{\infty}\!\frac{\tilde{\varepsilon}_{n}}{
\tilde{\varepsilon}_{n}^{2}+x^{2}\beta^{2}}\,{\rm e}^{-x^{2}/2}\,dx  \, .
\label{K00-2}
\end{eqnarray}
The equivalence between Eqs.\ (\ref{K00}) and (\ref{K00-2}) can also be
checked numerically.

\subsection{Expression of $\kappa_{\rm GL}$}

Near $T_{c}$
where $\beta\!\ll\! \tilde{\varepsilon}_{n}$ holds, 
we may choose $N_{\rm cut}\!=\! 1$ in Eqs.\
(\ref{R_N})-(\ref{KN'N2})
and expand the resulting expressions 
with respect to $\beta/\tilde{\varepsilon}_{n}$.
This yields
$\tilde{K}^{0}_{0}\!
\approx\!1/\tilde{\varepsilon}_{n}\!-\!\beta^{2}/\tilde{\varepsilon}_{n}^{3}$
and
$\tilde{K}^{0}_{1}\!\approx\!\beta/\tilde{\varepsilon}_{n}^{2}$,
so that Eq.\ (\ref{f(1)}) can be approximated by
$\tilde{f}^{(1)}_{0}\!\approx\!1/\varepsilon_{n}\!-\!
(\beta^{2}\varepsilon_{n}\!+\!\langle\beta^{2}\rangle/2\tau)/\varepsilon_{n}^{2}
\tilde{\varepsilon}_{n}^{2}$
and 
$\tilde{f}^{(1)}_{1}\!\approx\!
\beta/\varepsilon_{n}\tilde{\varepsilon}_{n}$.
Using these results in Eq.\ (\ref{S}) and retaining only terms of the leading order
in $\beta$, we obtain
\begin{subequations}
\label{S-GL}
\begin{eqnarray}
S_{2}=
\frac{1}{2d(\pi T_{c})^{2}}\sum_{n=0}^{\infty}
\frac{1}{(2n\!+\!1)^{2}(2n\!+\!1\!+\!1/2\pi\tau T_c)}
\, ,
\end{eqnarray}
\begin{eqnarray}
S_{4}=\frac{7\zeta(3)}{8(\pi T_{c})^{2}}\beta_{\rm A} \, ,
\end{eqnarray}
\begin{eqnarray}
S_{A}=S_{2}^{2}\beta_{\rm A}\, ,
\end{eqnarray}
\end{subequations}
where $d\!=\!2,3$ is the dimension of the system.
Substituting Eq.\ (\ref{S-GL}) into Eq.\ (\ref{H-B3}), 
we find the expression of $\kappa_{\rm GL}\!\equiv\! \kappa_{2}(T_{c})$ as
\begin{equation}
\kappa_{\rm GL}=\frac{d\,\pi T_{c}\sqrt{7\zeta(3)}/2}
{\displaystyle\sum_{n}[(2n\!+\!1)^{2}(2n\!+\!1\!+\!1/2\pi\tau T_c)]^{-1}}
\,\kappa_{0} \, . 
\label{kappaGL}
\end{equation}
This expression enables us to eliminate $\kappa_{0}$ in favor of $\kappa_{\rm GL}$.

The case with $p$-wave impurity scattering may be treated similarly by using
Eqs.\ (\ref{f(1)p2}) and (\ref{J(3)p2}) for $\tilde{f}^{(1)}_{N}$ and $J^{(3)}_{N}$,
respectively.
The resulting $\kappa_{\rm GL}$ is given by Eq.\ (\ref{kappaGL})
with a replacement of $\tau$ by the transport lifetime $\tau_{\rm tr}$
defined through
\begin{equation}
\frac{1}{\tau_{\rm tr}}\equiv \frac{1}{\tau}-\frac{1}{\tau_{1}}
 \, , 
\label{tauTr}
\end{equation}
in agreement with Eilenberger.\cite{Eilenberger67}

\subsection{\label{subsec:diffEilen}Eilenberger's results}

I now clarify the connection with Eilenberger's well-known results. \cite{Eilenberger67}
They are obtained by extracting from $I_{N_{1}N_{2}N_{3}N_{4}}^{(4)}$ 
of Eq.\ (\ref{I4})
a part which may be expressed in terms of $\beta_{\rm A}\!=\!1.16$ of Eq.\ (\ref{betaA}).

To see this, let us start from an alternative expression of
$I_{N_{1}N_{2}N_{3}N_{4}}^{(4)}$ 
for $N_{1}\!+\!N_{2}\!=\!N_{3}\!+\!N_{4}$:
\begin{eqnarray}
&&I_{N_{1}N_{2}N_{3}N_{4}}^{(4)}
=\sum_{N_{a}} \frac{V}{2}\sum_{\alpha=1}^{2} |\psi_{N_{a}{\bf 0}
\alpha}({\bf 0})|^{2}
\nonumber \\
&&\hspace{3mm}\times
\langle N_{1}N_{2}|
N_{1}\!+\!N_{2}\!-\!N_{a}N_{a}\rangle
\langle N_{3}N_{4}|
N_{1}\!+\!N_{2}\!-\!N_{a}N_{a}\rangle
\, . 
\nonumber \\
\label{I4-2}
\end{eqnarray}
Here $\psi_{N_{a}{\bf 0}\alpha}({\bf 0})$ and $\langle N_{1}N_{2}|N_{3}N_{4}\rangle$
are the quasiparticle wave function and the overlap integral
defined by Eqs.\ (3.12) and (3.23) of Ref.\ \onlinecite{Kita98-2}, respectively.
This identity can be proved by using Eqs.\ (3.22) and (3.24) of Ref.\ \onlinecite{Kita98-2}
and noting $\psi^{({\rm c})}_{N{\bf q}}({\bf r})\!=\!\psi^{({\rm r})}_{N{\bf q}}(2{\bf r})$
which both denote the present $\psi_{N{\bf q}}({\bf r})$.
If we retain only terms of $N_{a}\!=\!0$ in Eq.\ (\ref{I4-2}) and 
use $\frac{V}{2}\sum_{\alpha=1}^{2} |\psi_{0{\bf 0}\alpha}({\bf 0})|^{2}\!=\!\beta_{\rm A}$, 
we obtain an approximate expression for Eq.\ (\ref{I4-2}) as
\begin{eqnarray}
I_{N_{1}N_{2}N_{3}N_{4}}^{(4)}
&&\hspace{-3mm}\approx
\beta_{\rm A}\langle N_{1}N_{2}|
N_{1}\!+\!N_{2}0\rangle
\langle N_{3}N_{4}|
N_{1}\!+\!N_{2}0\rangle
\nonumber \\
&&\hspace{-3mm}=
\frac{(N_{1}\!+\!N_{2})!}{2^{N_{1}+N_{2}}\sqrt{N_{1}!N_{2}!N_{3}!N_{4}!}}\,\beta_{\rm A}
\nonumber \\
&&\hspace{-3mm}\equiv I_{N_{1}N_{2}N_{3}N_{4}}^{(4{\rm E})}
\, . 
\label{I4E}
\end{eqnarray}
Now, Eilenberger's approximation for $\kappa_{2}$ is given by Eqs.\ (2.7), (2.8), and (6.5)
of his paper\cite{Eilenberger67} and corresponds to
\begin{equation}
\kappa_{2}\approx(\kappa_{0}^{2}S_{4}^{({\rm E})}/2S_{2}^{2}\beta_{\rm A})^{1/2}
\equiv\kappa_{2}^{({\rm E})} \, ,
\label{kappa2E}
\end{equation}
in Eq.\ (\ref{H-B3}),
where $S_{4}^{({\rm E})}$ is obtained from Eq.\ (\ref{S4}) by replacing 
$I^{(4)}$ by $I^{(4{\rm E})}$.
Indeed, this procedure yields numerical agreements with his results.
As seen below in Sec.\ \ref{subsec:kappa2}, however, 
this approximation is not correct quantitatively far
beyond his estimation $\sim\! 1\%$.

It should also be noted that Eilenberger's definition
of $\kappa_{2}$ by Eq.\ (\ref{kappa2E})
is different from Maki's through Eq.\ (\ref{H-B3}) with respect to
\begin{equation}
\eta\equiv S_{A}/S_{2}^{2}\beta_{\rm A} \, .
\label{eta}
\end{equation}
I resume Maki's definition through Eq.\ (\ref{H-B3})
where $\kappa_{2}$ has a one-to-one
correspondence with the initial slope of the magnetization.
This is certainly more preferable than
expressing the slope with two parameters, $\kappa_{2}^{(E)}$ and $\eta$.

I finally comment on Eilenberger's analytic expression for $\eta$.
In addition to the approximation (\ref{I4E}),
it was obtained by integrating the Maxwell equation with a removal of a common
operator; see the argument above Eq.\ (6.4).\cite{Eilenberger67}
However, this procedure may bring an erroneous constant.
Indeed, his $B_{0}({\bf r})$ of Eq.\ (6.4)
does not satisfy the required condition $\int B_{0}({\bf r})\,d{\bf r}\!=\!0$.
Thus, his expression for $\eta$ is incorrect in the two respects
and cannot be obtained by adopting the approximation (\ref{I4E}) in Eq.\ (\ref{eta}).

\begin{figure}[t]
\includegraphics[width=0.75\linewidth]{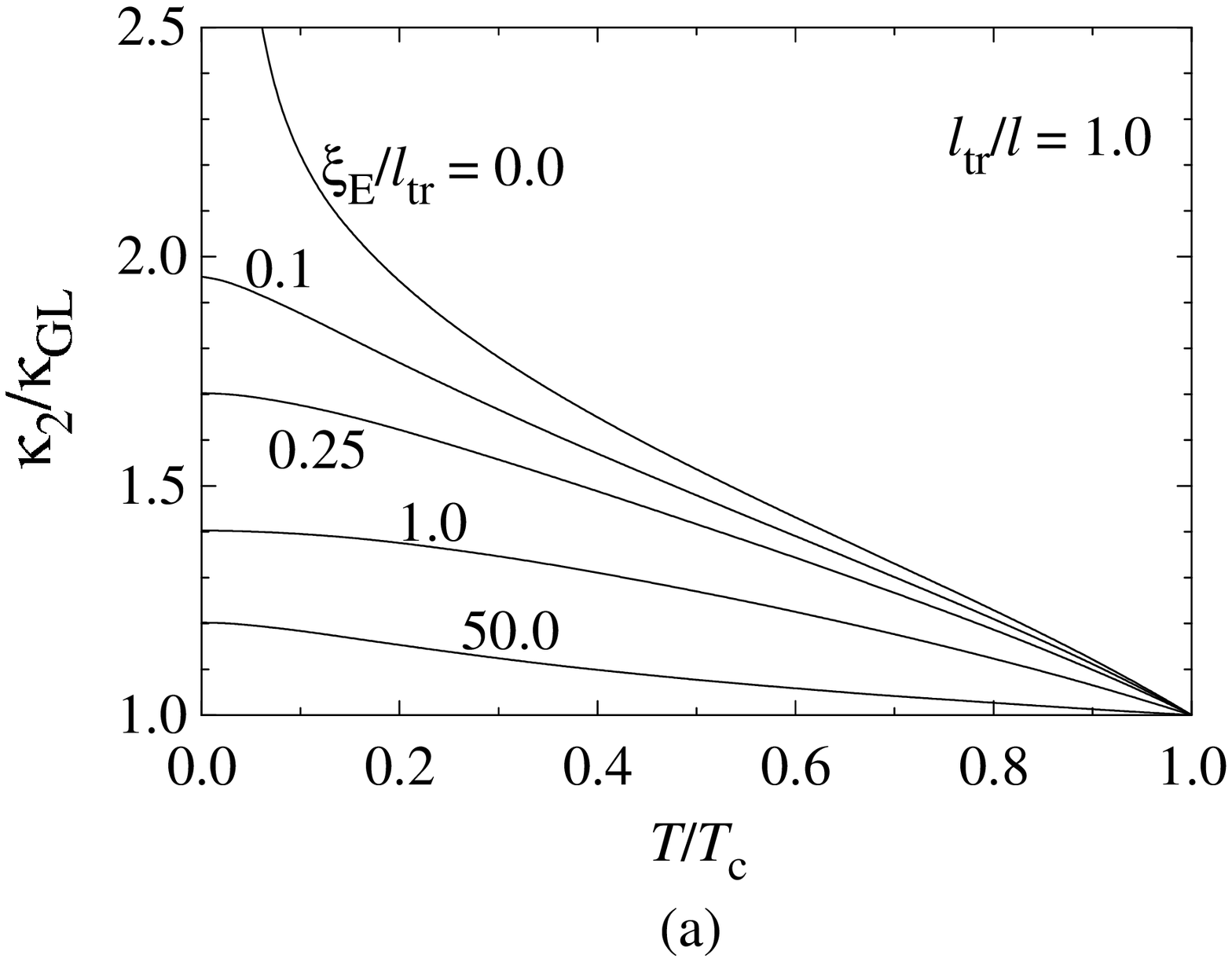}
 
\includegraphics[width=0.75\linewidth]{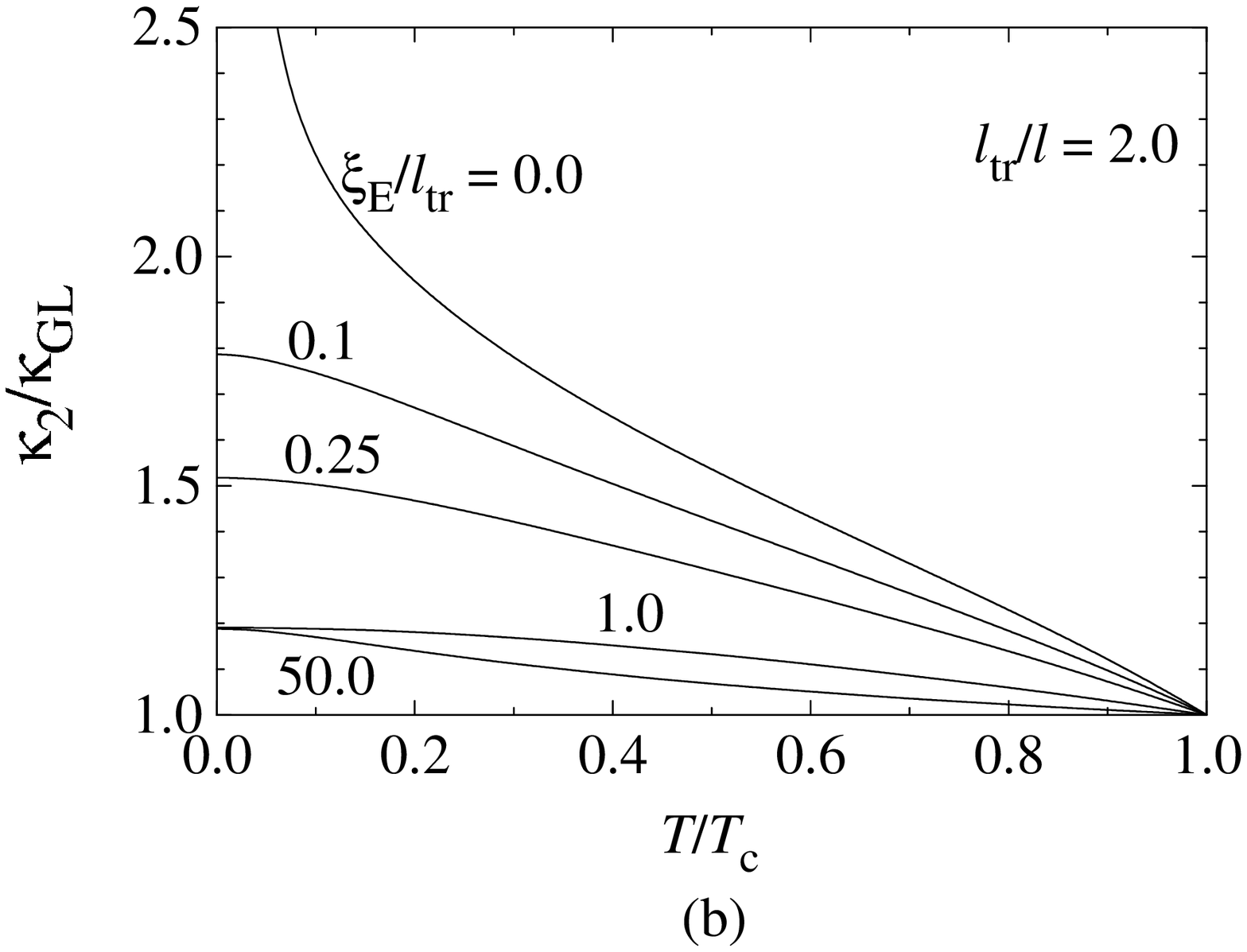} 
\caption{Temperature dependence of $\kappa_{2}/\kappa_{\rm GL}$ 
for several values of
$\xi_{\rm E}/l_{\rm tr}$
in the extreme type-II case $\kappa_{\rm GL}\!=\! 50$.
(a)  $l_{\rm tr}/l =1.0$; (b) $l_{\rm tr}/l =2.0$.}
\label{fig:1}
\end{figure}

\section{\label{sec:results}Numerical Results}

\subsection{\label{subsec:numproc}Numerical procedures}

I have adopted the same parameters
as those of Eilenberger:\cite{Eilenberger67}
\begin{equation}
\xi_{\rm E}/l_{\rm tr} \equiv 1/2\pi T_{c}\tau_{\rm tr} \, , \hspace{5mm}
l_{\rm tr}/l \equiv \tau_{\rm tr}/\tau\, ,
\end{equation}
to express different impurity concentrations.
Numerical calculations of Eqs.\ (\ref{pair3-1})-(\ref{S})
have been performed for each set of parameters by restricting every summation over 
the Matsubara frequencies for those satisfying 
$\varepsilon_{n}\!\leq\!\varepsilon_{c}$.
Choosing $\varepsilon_{c}\!=\!50\!\sim\!100$ has been sufficient to obtain
an accuracy of $\sim\!0.01\%$ for $\kappa_{2}$. 
On the other hand, summations over the Landau levels 
have been truncated at $N\!=\!N_{\rm cut}$ where I put ${\cal R}_{N_{\rm cut}}\!=\!1$
in the calculation of $K_{N}^{N'}$; see Sec.\ \ref{subsec:KN'N} for the details.
Enough convergence has been obtained by choosing
$N_{\rm cut}\!=\!4$, $40$, $100$, $200$, $500$, and $2000$ for 
$\xi_{\rm E}/l_{\rm tr}\!=\!50$, $1.0$, $0.5$, $0.1$, and $0.05$, respectively.
Finally, integrations over $\theta$ have been performed by
Simpson's formula with $N_{\rm cut}\!+\! 1$ integration
points for $0\!\leq\!\theta\!\leq\!\pi/2$.

\subsection{\label{subsec:kappa2}Results for $\kappa_{2}$}
Figure 1 shows $\kappa_{2}/\kappa_{\rm GL}$ as a function of $T/T_{c}$
for different impurity concentrations.
The upper one is for $
l_{\rm tr}/l \!=\! 1.0 $, i.e., the case without $p$-wave impurity scattering,
whereas the lower one is for $l_{\rm tr}/l \!=\! 2.0$.
They are calculated in an extreme type-II case of $\kappa_{\rm GL}\!=\! 50$,
so that they directly correspond to Eilenberger's
results for $l_{\rm tr}/l\!=\! 1$ and $2$, respectively.\cite{Eilenberger67}
These curves show qualitatively the same behaviors as those of Eilenberger's, 
including the divergence in the clean limit for $T\!\rightarrow\! 0$,
as predicted by Maki and Tsuzuki. \cite{MT65}
Except the curves in the dirty limit, however, 
marked quantitative differences are seen.
For example, $\kappa_{2}(T\!=\!0)/\kappa_{\rm GL}$ for $(\xi_{\rm E}/l_{\rm tr},l_{\rm tr}/l)\!=\! 
(1.0,1.0)$ is $1.40$ from the present calculation,
whereas it is $1.50$ from Eilenberger's.
Thus, we realize that Eilenberger's approximation (\ref{kappa2E}) yields quantitative errors
of $\alt 20\%$ for the deviation $\kappa_{2}/\kappa_{\rm GL}\!-\! 1$.
Comparing the two figures, we observe the followings:
(i) The results in the dirty limit are the same between 
$l_{\rm tr}/l \!=\! 1 $ and $2$;
(ii) $p$-wave scattering has a general tendency 
to lower the values of $\kappa_{2}$, and
also produces a non-monotonic behavior in
$\kappa_{2}/\kappa_{\rm GL}$ as a function 
of $\xi_{\rm E}/l_{\rm tr}$.

\begin{figure}[t]
\includegraphics[width=0.75\linewidth]{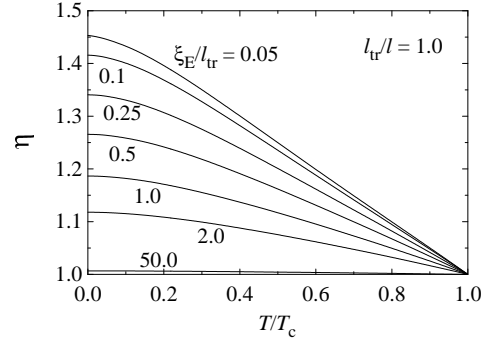}%
\caption{Temperature dependence of $\eta$ defined by Eq.\ (\ref{eta})
for several values of
$\xi_{\rm E}/l_{\rm tr}$ with $l_{\rm tr}/l\!=\! 1.0$.}
\label{fig:2}
\end{figure}
\begin{figure}[b]
\includegraphics[width=0.75\linewidth]{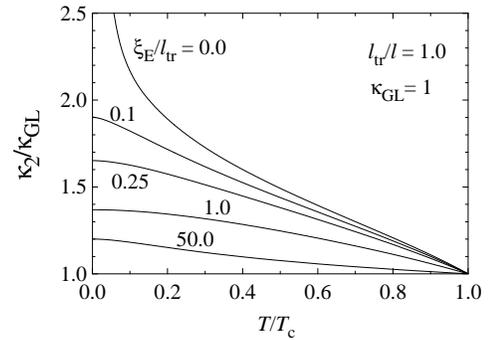}%
\caption{Temperature dependence of $\kappa_{2}/\kappa_{\rm GL}$ 
for $\kappa_{\rm GL}\!=\! 1$ with $l_{\rm tr}/l =1.0$.}
\label{fig:3}
\end{figure}

Figure 2 displays $\eta$ defined by Eq.\ (\ref{eta})
as a function of $T/T_{c}$ for
$\xi_{\rm E}/l_{\rm tr}\!=\!0.05$-$50.0$ and $l_{\rm tr}/l\!=\! 1.0$.
This quantity becomes relevant for small values of $\kappa_{\rm GL}$
at low temperatures, as may be realized by Eq.\ (\ref{H-B3}).
The curves also deviate substantially from Eilenberger's results. 
For example, $\eta$ for $\xi_{\rm E}/l_{\rm tr}\!=\!0.25$ at $T\!=\!0$ is $1.34$ 
from the present calculation, whereas it is $\sim\!1.11$ from Eilenberger's.
Generally, the values are larger than those of Eilenberger's.
This fact implies that $\kappa_{2}(T)$ for $\kappa_{\rm GL}\!\sim\! 1$ 
becomes smaller than the evaluation of Eilenberger.

To see the dependence of $\kappa_{2}/\kappa_{\rm GL}$ on $\kappa_{\rm GL}$ explicitly,
I have performed a calculation of $\kappa_{2}$
near the type-I-type-II boundary of $\kappa_{\rm GL}\!=\! 1.0$.
Figure 3 plots the results for
$\xi_{\rm E}/l_{\rm tr}\!=\!0.0$-$50.0$ and $l_{\rm tr}/l\!=\! 1$.
Compared with Fig.\ 1(a), we observe that each curve is slightly shifted downward.
However, the changes are surprisingly small, considering the closeness to the 
type-I-type-II boundary.
We thus realize that the factor $S_{A}/S_{2}^{2}\!=\!\eta\beta_{\rm A}$ 
in Eq.\ (\ref{H-B3}) can be neglected practically for $\kappa_{\rm GL}\!\agt\! 5$,
as already observed by Eilenberger.\cite{Eilenberger67}

\begin{figure}[t]
\includegraphics[width=0.75\linewidth]{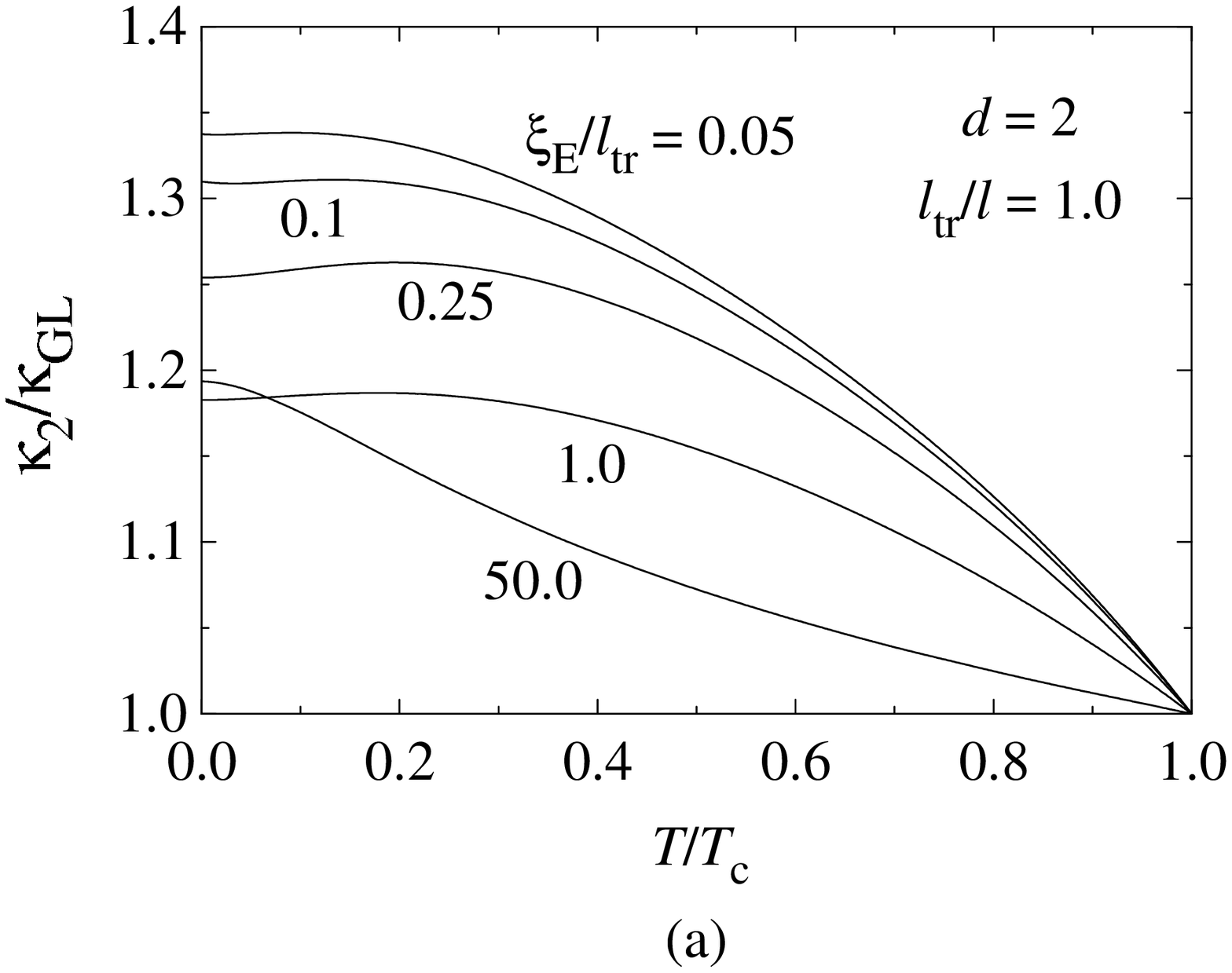}
 
\includegraphics[width=0.75\linewidth]{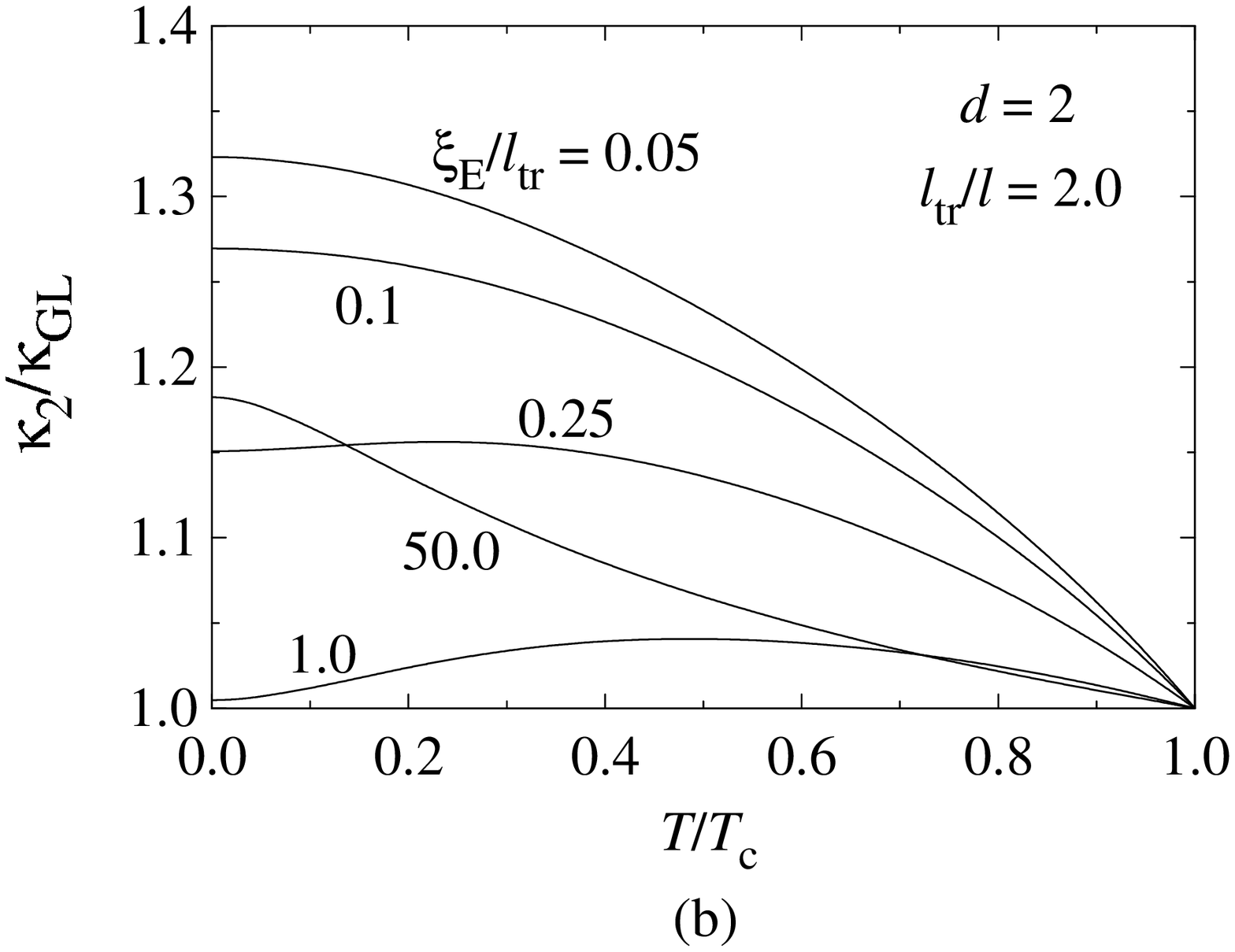} 
\caption{Temperature dependence of $\kappa_{2}/\kappa_{\rm GL}$ 
for an isotropic two-dimensional system
in the extreme type-II case $\kappa_{\rm GL}\!=\! 50$.
(a)  $l_{\rm tr}/l =1.0$; (b) $l_{\rm tr}/l =2.0$.}
\label{fig:4}
\end{figure}

The above calculations are performed for an idealized spherical Fermi surface.
However, real superconductors are often characterized by complicated Fermi surfaces.
To see the dependence of $\kappa_{2}/\kappa_{\rm GL}$ on Fermi-surface structures,
I have performed an isotropic two-dimensional calculation described in Sec.\ \ref{subsec:2d}.
Figure 4 shows the results, where the parameters are the same as those in Fig.\ 1.
The curves for $\xi_{\rm E}/l_{\rm tr}\!=\!50$ 
are almost the same as those in Fig.\ 1.
Thus, in the dirty limit, we have a universal curve which depends neither on
detailed Fermi-surface structures nor fine features of the impurity scattering.
As the system becomes cleaner, however, differences due to the two factors 
emerge eventually.
In fact, we observe that each curve for $\xi_{\rm E}/l_{\rm tr}\!\alt\!1.0$ 
in Fig.\ 4
deviates far less from $1$ than the corresponding one in Fig.\ 1, and 
the temperature dependence is also weaker.
Another point to be mentioned is that, even for $\xi_{\rm E}/l_{\rm tr}\!=\!0.05$,
we see no trace of divergence as $T\!\rightarrow\! 0$.
Indeed, a closer examination of the analytic results by Maki and Tsuzuki\cite{MT65} and 
Eilenberger\cite{Eilenberger67}
enables us to realize that it is the region $\theta\!\sim\! 0$ in three dimensions 
which is responsible for the divergence of $\kappa_{2}$.
Thus, we may conclude that $\kappa_{2}$ in two dimensions
remains finite even in the clean limit as $T\!\rightarrow\! 0$.
In general, $\kappa_{2}$ will remain finite if the relevant Fermi surface does not close along
the direction of the magnetic field.

\begin{figure}[t]
\includegraphics[width=0.75\linewidth]{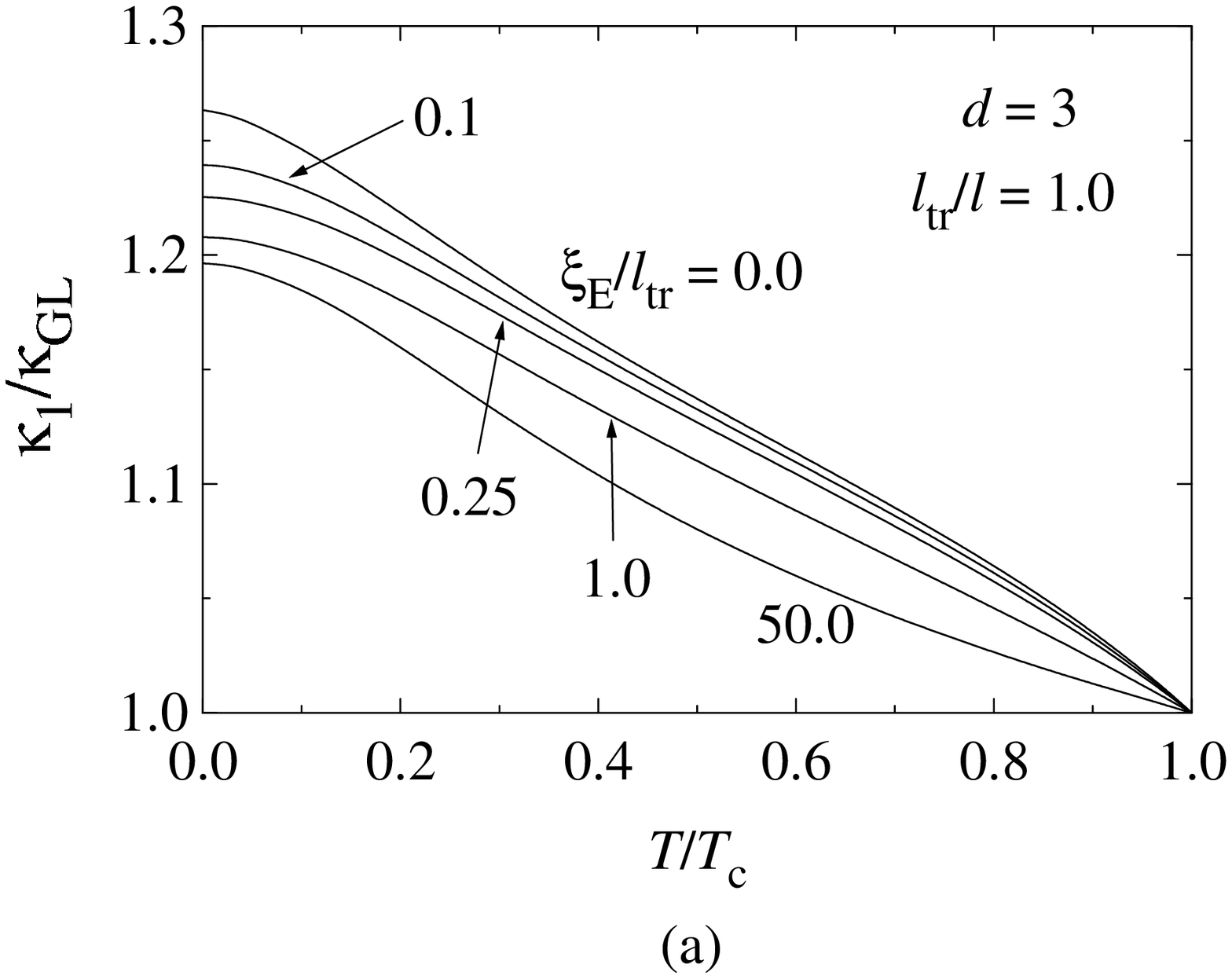}
 
\includegraphics[width=0.75\linewidth]{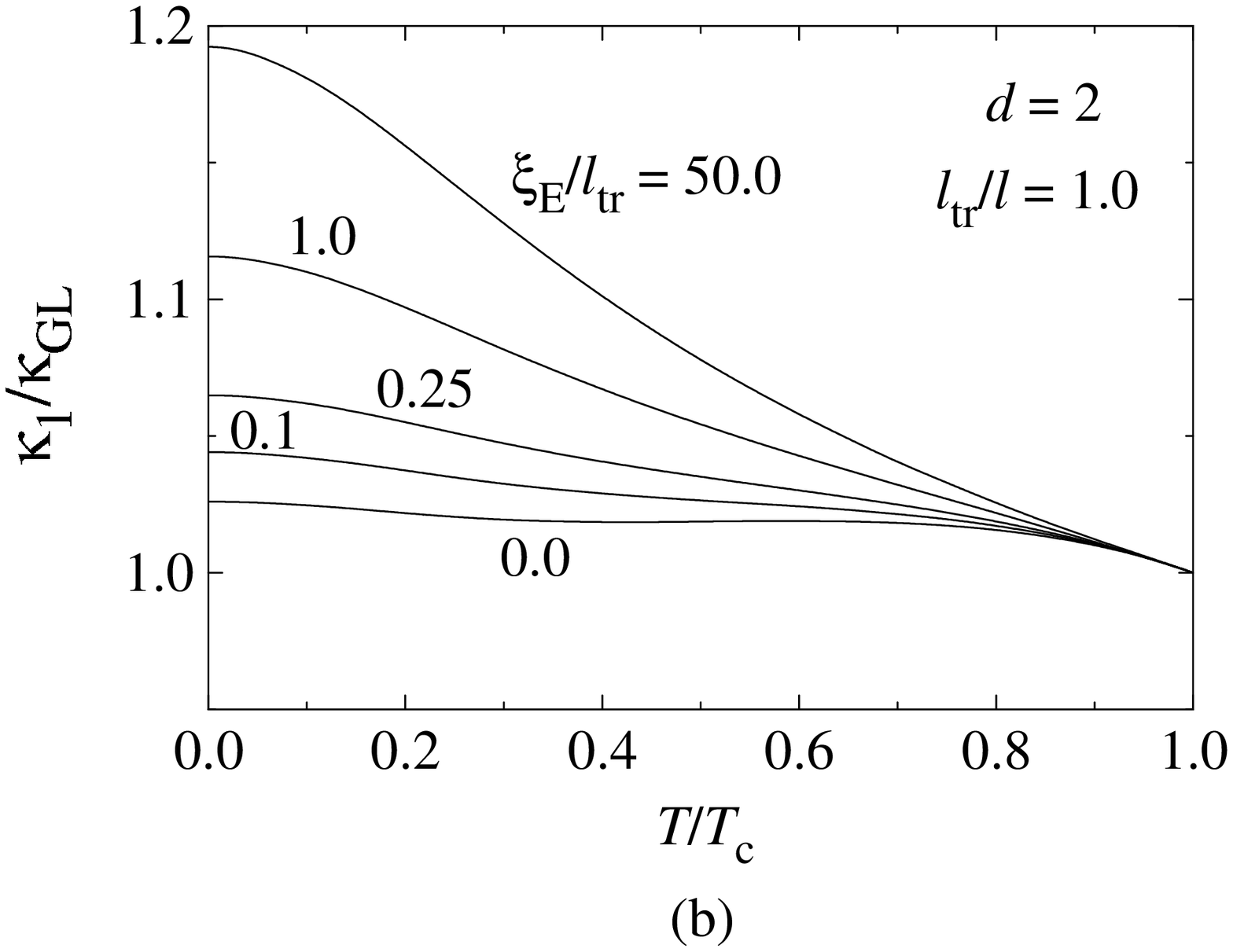} 
\caption{Temperature dependence of $\kappa_{1}/\kappa_{\rm GL}$ 
for several values of
$\xi_{\rm E}/l_{\rm tr}$ with $l_{\rm tr}/l =1.0$.
(a) $d\!=\!3$ ; (b) $d\!=\!2$.}
\label{fig:5}
\end{figure}

\subsection{\label{subsec:kappa1}Results for $\kappa_{1}$}
The Maki parameter $\kappa_{1}$ is defined by\cite{Maki64}
\begin{equation}
\kappa_{1}\equiv{H_{c2}}/{\sqrt{2}H_{c}} \, ,
\end{equation}
where $H_{c}\!=\!H_{c}(T)$ is the thermodynamic critical field.
The preceding results for $\kappa_{2}$ suggest that
$\kappa_{1}(T)/\kappa_{\rm GL}$ may also exhibit considerable dependence on
detailed Fermi-surface structures.

Figure 5 compares $\kappa_{1}(T)/\kappa_{\rm GL}$ between
two and three dimensions for $l_{\rm tr}/l\!=\!1.0$.
The curves for $\xi_{\rm E}/l_{\rm tr}\!=\!50$ 
show almost the same behavior.
As $\xi_{\rm E}/l_{\rm tr}$ becomes smaller, however,
the two cases display a marked difference.
Indeed, $\kappa_{1}(T)/\kappa_{\rm GL}$ is seen to increase (decrease) 
in three (two) dimensions as $\xi_{\rm E}/l_{\rm tr}\!\rightarrow\! 0$.

\begin{figure}[t]
\includegraphics[width=0.75\linewidth]{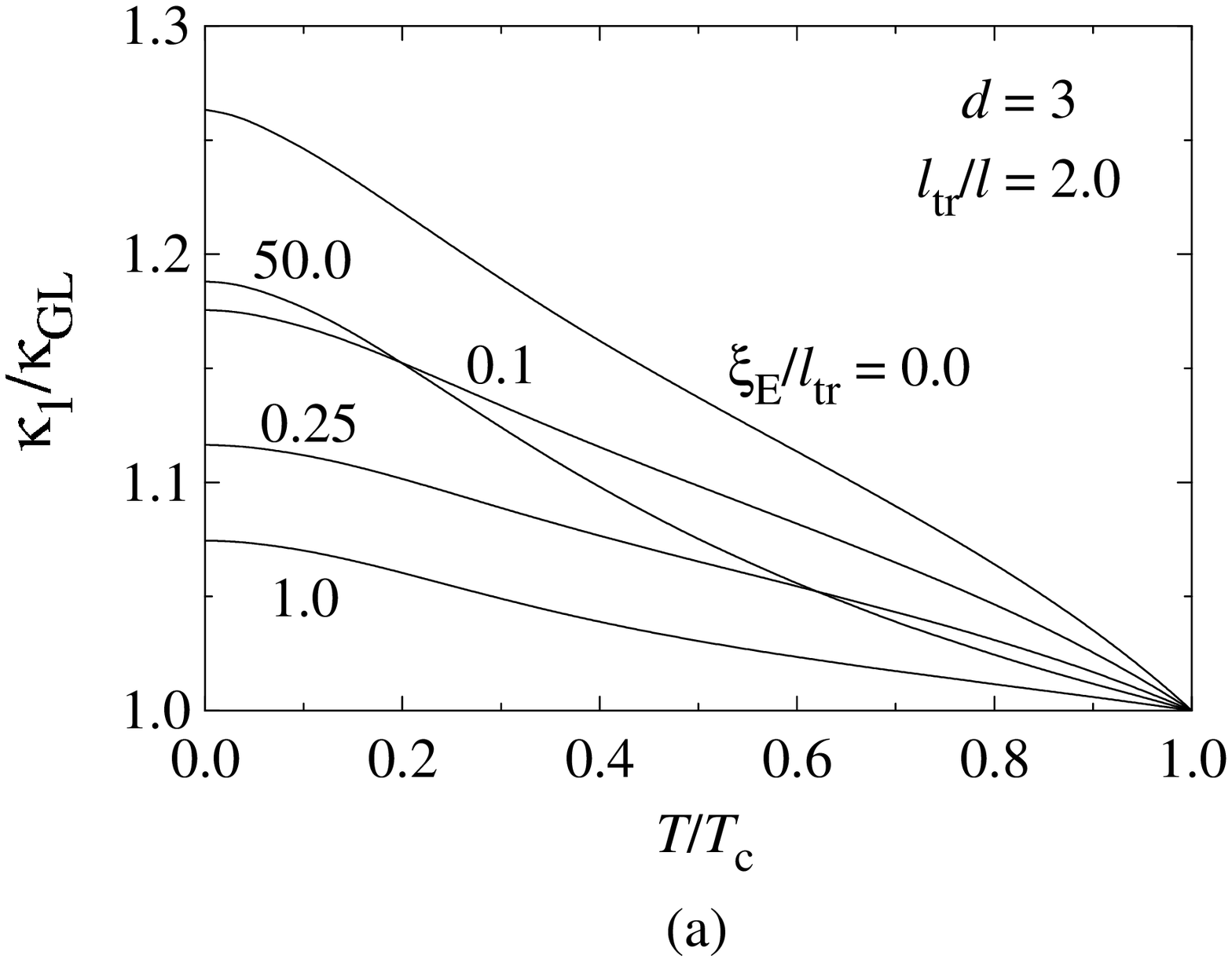}
 
\includegraphics[width=0.75\linewidth]{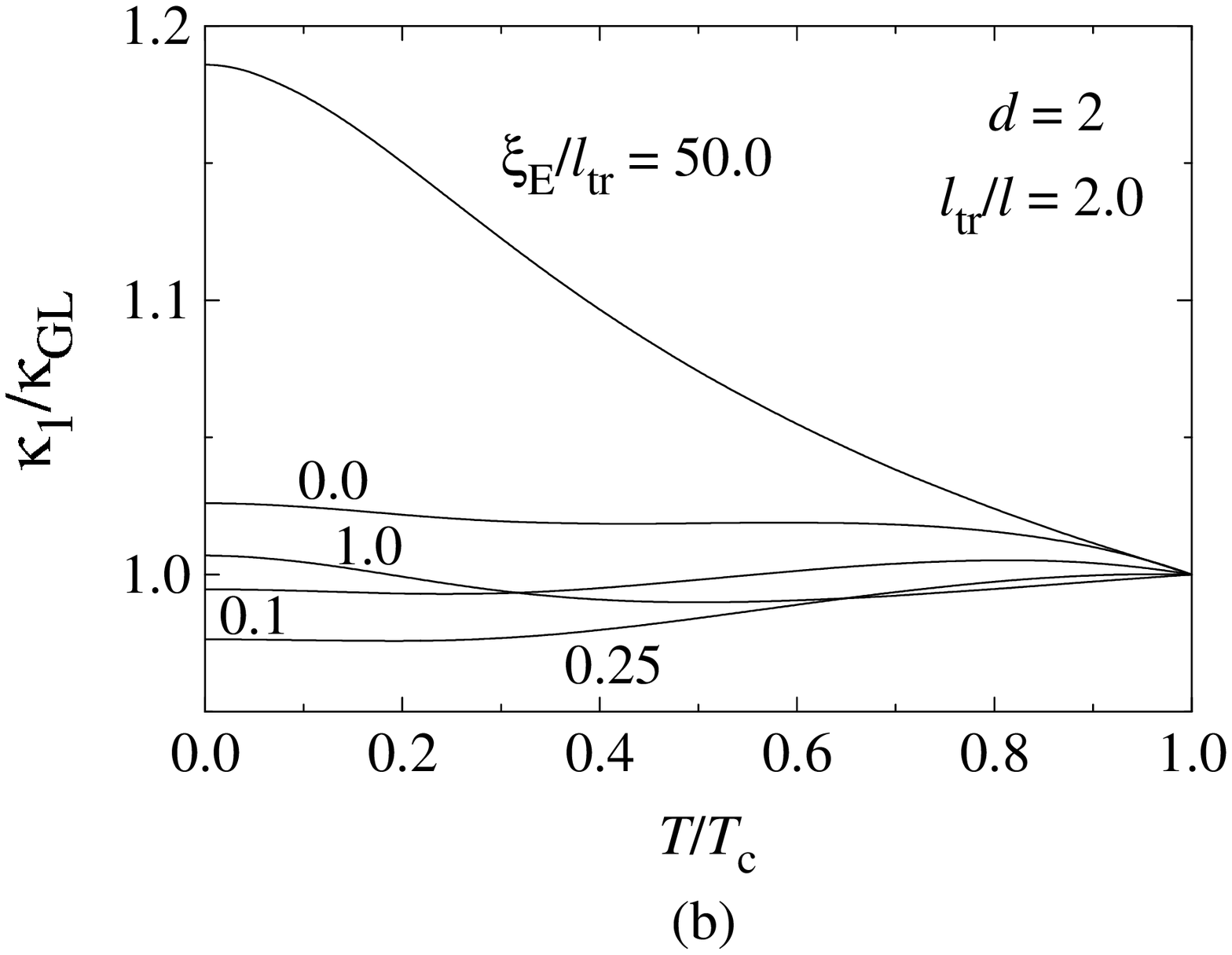} 
\caption{Temperature dependence of $\kappa_{1}/\kappa_{\rm GL}$ 
for several values of
$\xi_{\rm E}/l_{\rm tr}$ with $l_{\rm tr}/l =2.0$.
(a) $d\!=\!3$ ; (b) $d\!=\!2$.}
\label{fig:6}
\end{figure}

Figure 6 shows curves of $\kappa_{1}(T)/\kappa_{\rm GL}$ in
two and three dimensions for $l_{\rm tr}/l\!=\!2.0$.
Again the $p$-wave impurity scattering is seen to lower the value of $\kappa_{1}/\kappa_{\rm GL}$,
and also introduces non-monotonicity in $\kappa_{1}/\kappa_{\rm GL}$
as a function of $\xi_{\rm E}/l_{\rm tr}$. 
Especially in two dimensions for $\xi_{\rm E}/l_{\rm tr}\!=\!0.1$-$1.0$,
$\kappa_{1}/\kappa_{\rm GL}$ becomes smaller than $1$ over finite temperature ranges, i.e.,
the emperical inequality $\kappa_{2}\!\geq\!\kappa_{1}\!\geq\!\kappa_{\rm GL}$
is not satisfied here, even without spin paramagnetism.\cite{Maki66}

A substantial dependence of $H_{c2}$ on Fermi-surface structures may be 
realized more clearly by looking at the temperature dependence of the
reduced critical field introduced by Helfand and Werthamer:\cite{HW66}
\begin{eqnarray}
h^{*}(t)\equiv \frac{H_{c2}(t)}{-dH_{c2}(t)/dt|_{t=1}} \, ,
\end{eqnarray}
where $t\!\equiv T/T_{c}$.
Figure 7 compares $h^{*}(t)$
between two and three dimensions
for both the clean and dirty limits.
The curves coincide in the dirty limit, whereas 
those in the clean limit show a marked quantitative difference.
We also observe that $h^{*}(t)$ in
two dimensions is a rather sensitive function of purity.
A considerable reduction of $h^{*}_{d=2}(t)$ in the pure limit 
from $h^{*}_{d=3}(t)$
may be attributed to the pair breaking by supercurrent.
This effect is more effective in two dimensions.
Indeed, a point on the cylindrical Fermi surface is equivalent
to a point on the equator of the spherical Fermi surface perpendicular to ${\bf H}$
where the pair breaking is most effective.
This fact can be seen clearly in the polar-angle dependence of 
the density of states calculated by 
Brandt, Pesch, and Tewordt.\cite{BPT67}
Put it another way, if the relevant Fermi surface does not have a closed
orbit perpendicular to ${\bf H}$,
the corresponding $h^{*}(t)$ in the clean limit
will be enhanced over the prediction for the spherical 
Fermi surface.

A considerable reduction of $h^{*}(t)$ or $\kappa_{1}(t)$
in the presence of spin paramagnetism
was established by Werthamer, Helfand, and Hohenberg,\cite{WHH66}
and also by Maki.\cite{Maki66}
The present results indicate unambiguously that
the Fermi-surface structure is also an important factor for $h^{*}(t)$ 
in clean systems, as already noticed by Helfand and Werthamer,\cite{HW66}
Hohenberg and Werthamer,\cite{HW67} and
Werthamer and McMillan.\cite{WM67}

\begin{figure}[t]
\includegraphics[width=0.75\linewidth]{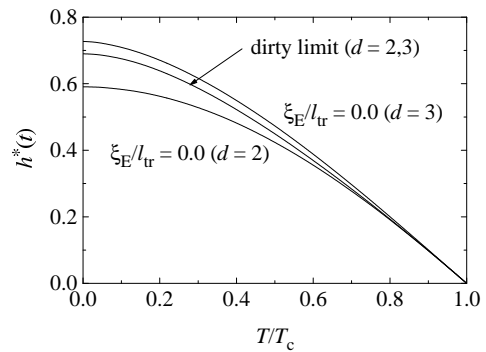}
\caption{Temperature dependence of the reduced critical field $h^{*}(t)$ 
for the dirty limit of $d\!=\!2,3$ (mid curve) and the clean limit 
of $d\!=\!2$  (lower curve) and  $d\!=\!3$  (upper curve).}
\label{fig:7}
\end{figure}

\subsection{\label{subsec:Delta0}Results for the pair potential}
\begin{figure}[t]
\includegraphics[width=0.75\linewidth]{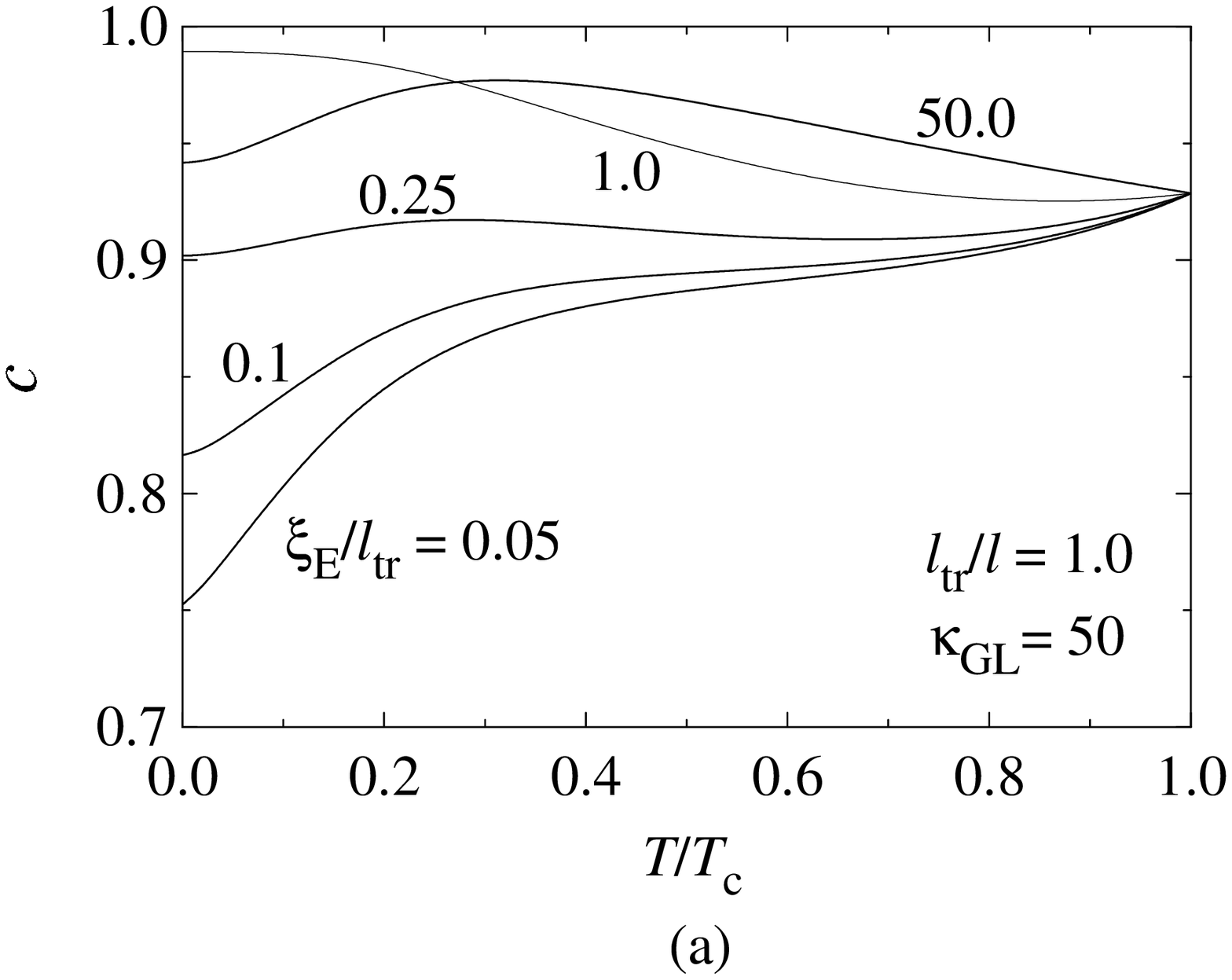}
 
\includegraphics[width=0.75\linewidth]{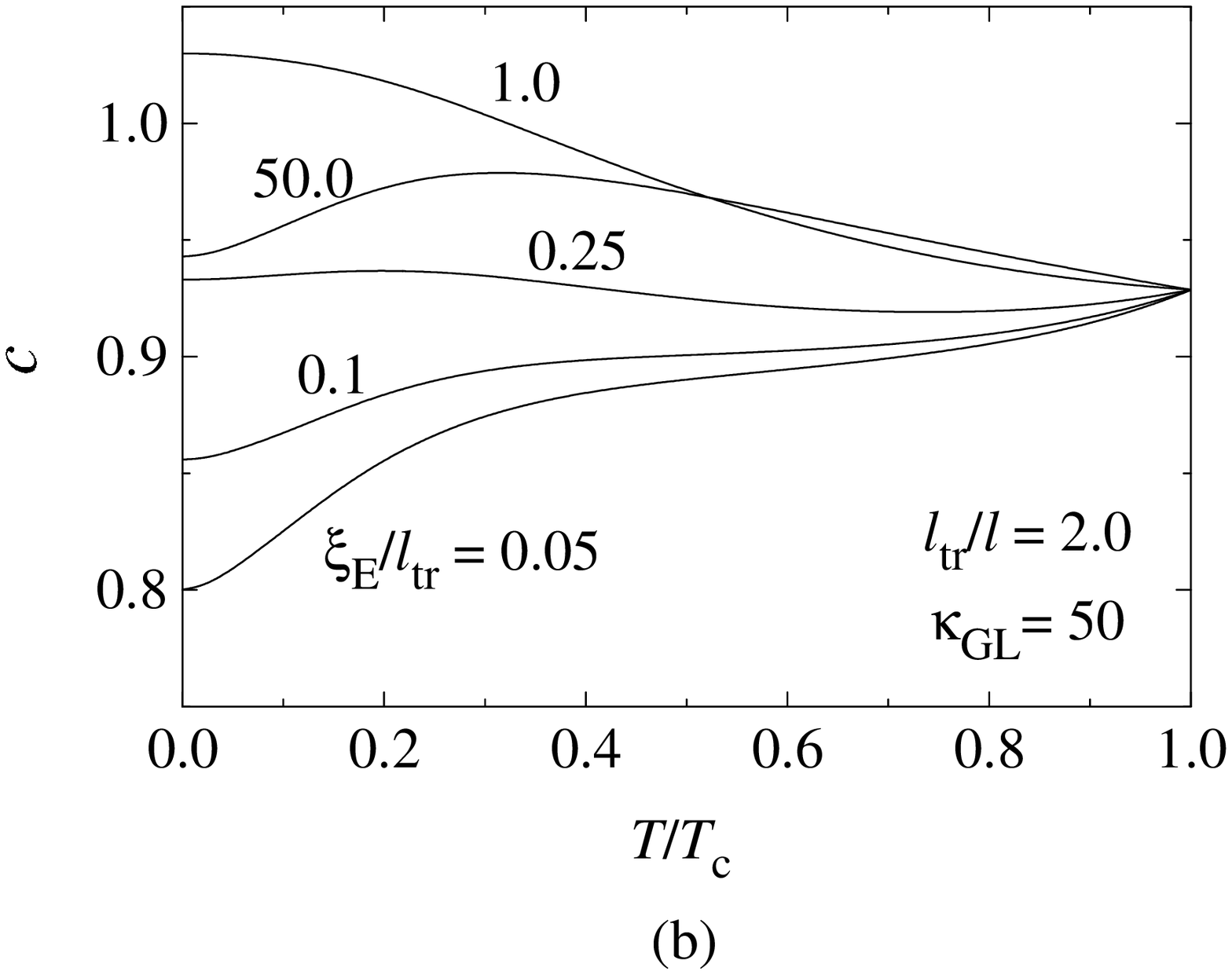}
\caption{The coefficient $c(T)\!\equiv\!(1\!-\!B/H_{c2})^{1/2}\Delta(T)/\Delta_{0}(B,T)$ in
the extreme type-II case $\kappa_{\rm GL}\!=\! 50$
as a function of $T/T_{c}$ for several
values of $\xi_{\rm E}/l_{\rm tr}$.
(a) $l_{\rm tr}/l =1.0$ ; (b) $l_{\rm tr}/l =2.0$.}
\label{fig:8}
\end{figure}

A quantity of fundamental importance is
the coefficient $\Delta_{0}$,
which is equal to 
the spatial average $\sqrt{\langle |\Delta({\bf r})|^{2}\rangle}$ of the pair potential
and relevant to all thermodynamic and transport properties near $H_{c2}$.
It is physically more meaningful to express it as a function of
the real average field $B$ in the bulk instead of $H$. 
Equation (\ref{pair3-3b}) shows that $\Delta_{0}(B)$ 
is proportional to $(H_{c2}\!-\!B)^{1/2}$ near $H_{c2}$.
I here express this $\Delta_{0}$ by using the energy gap $\Delta(T)$ at $B\!=\!0$ as
\begin{equation}
\Delta_{0}(B,T)= c(T) (1\!-\!B/H_{c2})^{1/2}\Delta(T)\, .
\end{equation}
Then the coefficient $c(T)$ should be of the order of $1$.

Figure 8, calculated for the spherical Fermi surface, 
displays temperature dependence of $c(T)$ 
in an extreme type-II case of $\kappa_{\rm GL}\!=\! 50$
for (a) $l_{\rm tr}/l\!=\!1.0$ and (b) $l_{\rm tr}/l\!=\!2.0$.
Thus $c(T)\!\sim\!1$, as expected,
having the same value $0.929$ at $T_{c}$.
Differences among different $\xi_{\rm E}/l_{\rm tr}$ grow at lower temperatures,
and $c(T)$ for $\xi_{\rm E}/l_{\rm tr}\!\alt\! 0.1$ drops rapidly near $T\!=\!0$. 
Indeed, $c(T)$ in the clean limit for three dimensions 
is expected to reach $0$ as $T\!\rightarrow\!0$,
corresponding to the divergence of $\kappa_{2}$.
This also implies that
the expansion in $\Delta({\bf r})$ near $H_{c2}$ is no longer valid in this limit.\cite{FH69}
The curves in the dirty limit are the same between 
$l_{\rm tr}/l\!=\!1.0$ and $l_{\rm tr}/l\!=\!2.0$.
For $\xi_{\rm E}/l_{\rm tr}\!\alt\! 1.0$, however,
each curve for $l_{\rm tr}/l\!=\!2.0$ at low temperatures
has larger values than the corresponding one
for $l_{\rm tr}/l\!=\!1.0$.
Thus, finite $p$-wave scattering
in clean systems tends to increase $c(T)$.

\begin{figure}[b]
\includegraphics[width=0.75\linewidth]{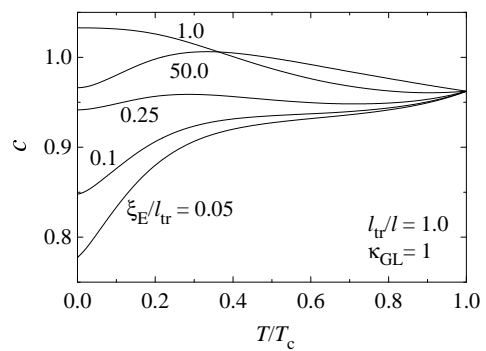}
\caption{The coefficient $c(T)$ for 
$\kappa_{\rm GL}\!=\! 1$
as a function of $T/T_{c}$ for several
values of $\xi_{\rm E}/l_{\rm tr}$
with $l_{\rm tr}/l =1.0$.}
\label{fig:9}
\end{figure}

The coefficient $c(T)$ also increases mildly as $\kappa_{\rm GL}$ becomes smaller,
as realized by comparing Fig.\ 9 for $\kappa_{\rm GL}\!=\! 1$
with Fig.\ 8(a) for $\kappa_{\rm GL}\!=\! 50$.

Figure 10 plots results of the two-dimensional calculations
performed with the same parameters as those in Fig.\ 8.
The curves for the dirty limit are the same between two and three dimensions.
As the system becomes cleaner, however, the coefficient $c(T)$ for two dimensions 
becomes larger than the corresponding one for three dimensions.
Thus, for clean systems, we observe once again a considerable dependence of the coefficient $c(T)$
on Fermi-surface structures.

\begin{figure}[t]
\includegraphics[width=0.75\linewidth]{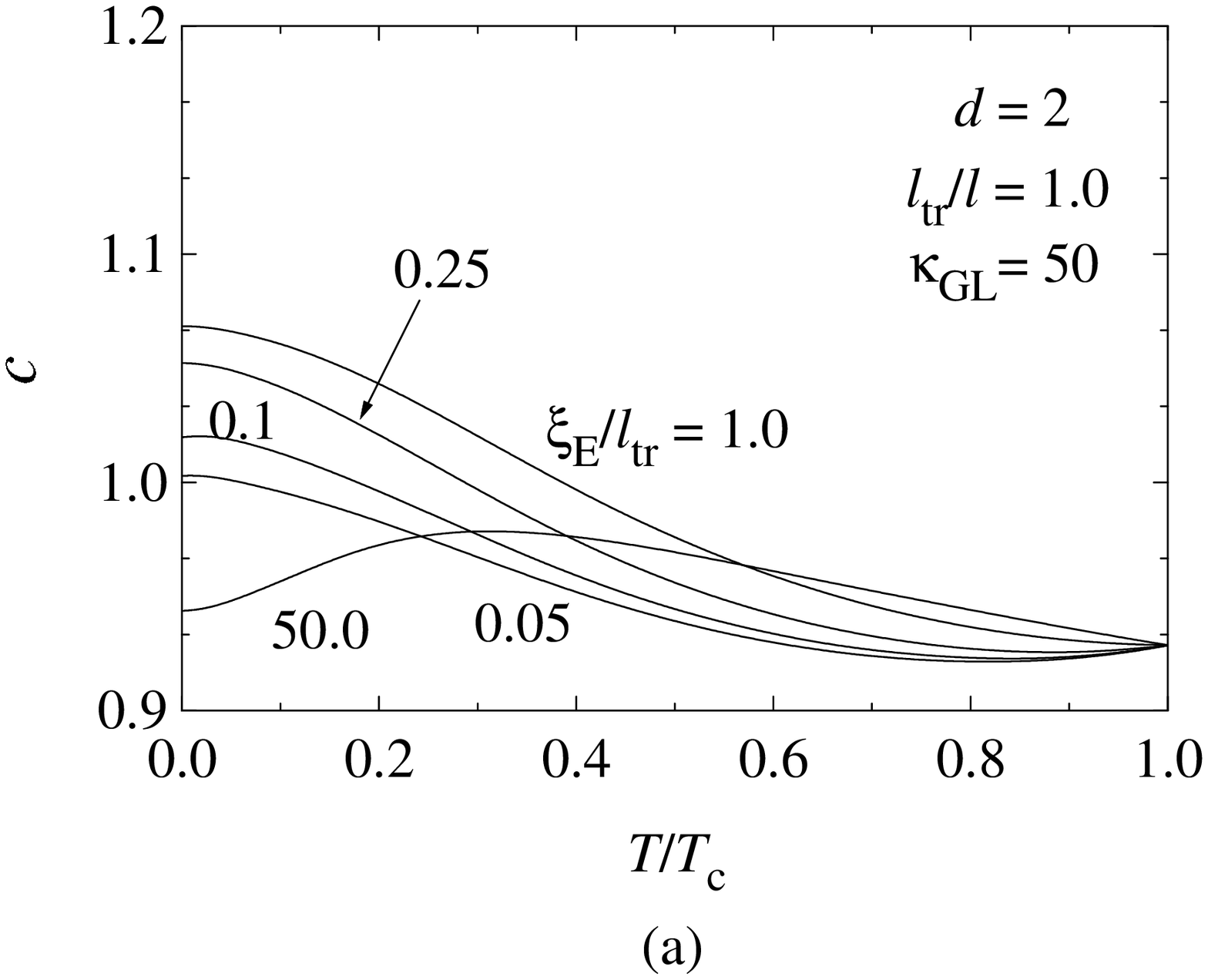}
 
\includegraphics[width=0.75\linewidth]{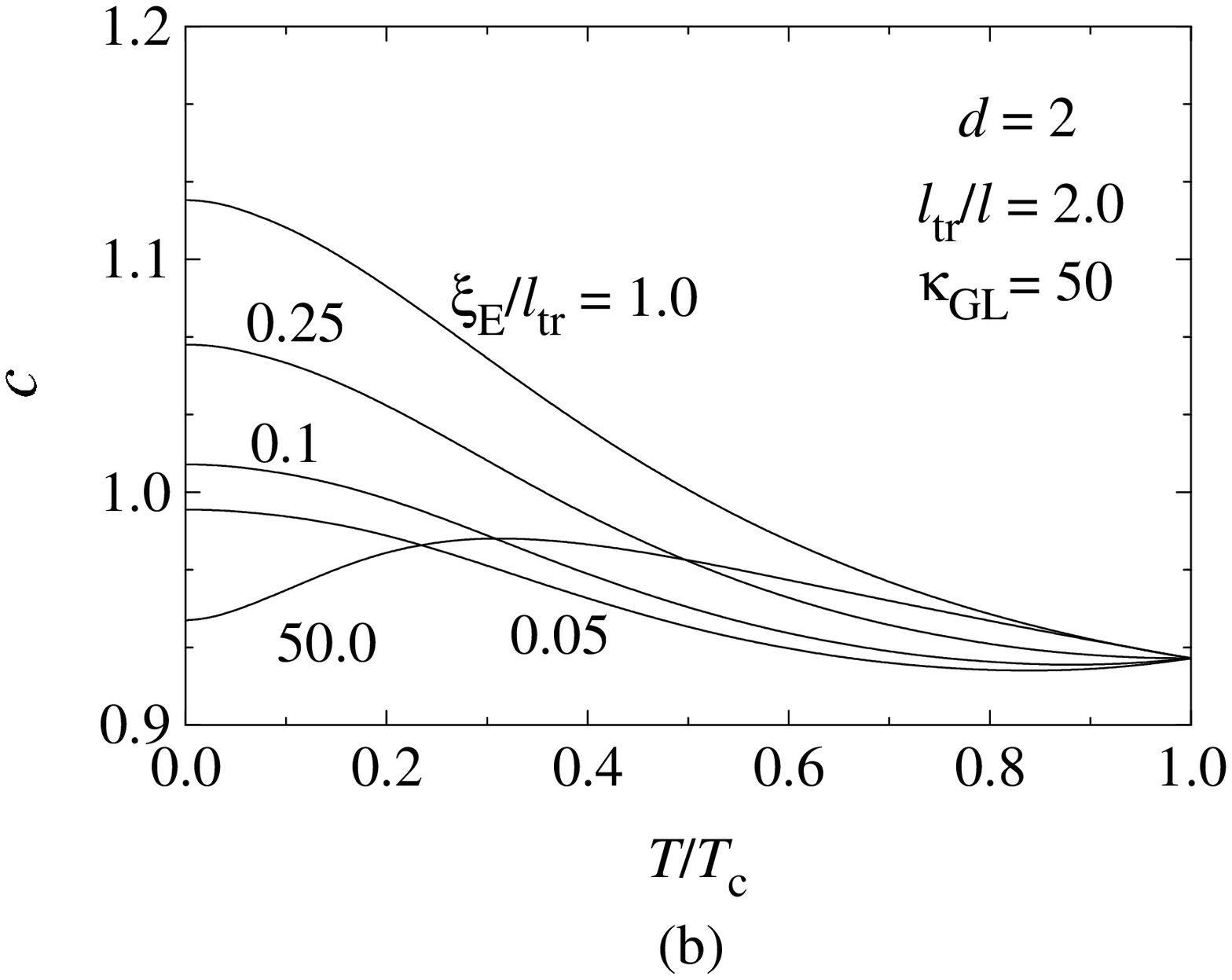} 
\caption{The coefficient $c(T)$ 
in two dimensions with 
$\kappa_{\rm GL}\!=\! 50$
as a function of $T/T_{c}$ for several
values of $\xi_{\rm E}/l_{\rm tr}$.
(a) $l_{\rm tr}/l =1.0$ ; (b) $l_{\rm tr}/l =2.0$.}
\label{fig:10}
\end{figure}
\section{\label{sec:summary}summary}

This paper has presented revised calculations of the Maki parameters $\kappa_{1}$ and
$\kappa_{2}$ as well as the spatial average $\langle |\Delta({\bf r})|^{2}\rangle$ near $H_{c2}$.
Eilenberger's results for $\kappa_{2}$ have been corrected appropriately,
as described in Sec.\ \ref{subsec:diffEilen}.
The analytic expressions derived in Secs.\ \ref{subsec:solutions} and \ref{subsec:KN'N}
have been useful to carry out efficient calculations for 
both two and three dimensions with isotropic Fermi surfaces
and arbitrary impurity concentrations.
Thereby found are large quantitative differences of the parameters 
between two and three dimensions
(except in the dirty limit where there are no differences between the two cases).
For example, no trace of divergence in $\kappa_{2}(T\!\rightarrow\! 0)$ is found
for the clean limit in two dimensions.

The present results clearly indicates 
the necessity of considering detailed Fermi-surface structures from 
first-principle calculations for an quantitative understanding of the Maki parameters
in clean superconductors.
This was already recognized by Helfand and Werthamer,\cite{HW66}
by Hohenberg and Werthamer,\cite{HW67} and also by Werthamer and McMillan\cite{WM67} 
when their strong-coupling calculation could not explain
a large deviation of $\kappa_{1}/\kappa_{\rm GL}$ 
observed in pure niobium\cite{MS65,FSS66} 
and vanadium\cite{RK66} from the theoretical prediction of Helfand and Werthamer.\cite{HW66}
Efforts have been made along this line to establish a realistic calculation 
of $\kappa_{1}$, or equivalently, 
$H_{c2}$.\cite{HW67,TN70,Mattheiss70,Teichler75,EP76,Butler80,YK80,PS87,RS90,Weber91,Langmann92}
However, little progress seems to have been
achieved with respect to $\kappa_{2}$.

The method developed here for $\kappa_{1}$ and $\kappa_{2}$
may be extended easily to include Fermi-surface structures and
anisotropic pairings. 
Some of the necessary modifications are: (i) to use the general expansion (\ref{DeltaExpand})
with $N\!=\!{\rm even}$ for the pair potential, rather than Eq.\ (\ref{DeltaExpand2});
(ii) to use more convenient basisfunctions 
than ${\rm e}^{im\varphi}$ in Eq.\ (\ref{fExpand}) for describing the 
${\bf k}_{\rm F}$ dependence of 
$f(\varepsilon_{n},{\bf k}_{\rm F},{\bf r})$, such as the Fermi-surface 
harmonics of Allen.\cite{Allen76,BA76,Allen78,AM82}
The corresponding matrix $\tilde{\cal M}$ in Eqs.\ (\ref{Eilen2-1}) and (\ref{Eilen2-3}) 
is no longer tridiagonal, but may be inverted rather easily
with present high-speed computers.

\begin{acknowledgments}
I am grateful to M. Endres and D. Rainer for discussions about
free-energy functionals of the quasiclassical theory.
This research is supported by Grant-in-Aid for Scientific Research 
from the Ministry of Education, Culture, Sports, Science, and Technology
of Japan.
\end{acknowledgments}

\appendix

\section{\label{sec:H-B}Derivation of Eq.\ (\ref{H-B})}

To obtain Eq.\ (\ref{H-B}), let us start from
Eilenberger's free-energy functional\cite{Eilenberger68} per unit volume with 
$B$ chosen as an independent variable instead of $H$.
It is given in units of $N(0)\Delta(0)^{2}$ as
\begin{eqnarray}
\frac{F(B)}{V}
=\frac{1}{V}\int\!d{\bf r}
\biggl\{ \frac{\kappa_{0}^{2}}{2}\left[B^{2}\!+\!
({\bm \nabla}\!\times\!\tilde{\bf A})^{2}\right]+|\Delta({\bf r})|^{2}\ln\frac{T}{T_{c}}
\nonumber \\
+2\pi T\sum_{n=0}^{\infty}\left[\frac{|\Delta({\bf r})|^{2}}{\varepsilon_{n}}-
\langle I(\varepsilon_{n},{\bf k}_{\rm F},{\bf r})\rangle \right]
\biggr\} \, ,\,\,\,\,\,\,\,\,\,\,
\label{F}
\end{eqnarray}
where $I$ is defined by
\begin{eqnarray}
&&\hspace{-3mm}I\equiv \Delta^{\! *}f\!+\!\Delta f^{\dagger} +2\varepsilon_{n}(g\!-\!1)
+\frac{f\langle f^{\dagger}\rangle\!+\!\langle f\rangle f^{\dagger}}{4\tau}
+\frac{g\langle g\rangle\!-\! 1}{2\tau}
\nonumber \\
&&\hspace{3mm}+(g\!-\!1)\hat{\bf v}\cdot
\frac{f^{\dagger}({\bm \nabla}\!-\!i{\bf A})f-f({\bm \nabla}\!+\! i{\bf A})f^{\dagger}}
{2f f^{\dagger}} \, .
\label{I}
\end{eqnarray}
The functional derivatives of Eq.\ (\ref{F}) with respect to $f^{\dagger}$, $\Delta$, 
and $\bf A$ lead to Eqs.\ (\ref{Eilen})-(\ref{Maxwell}), respectively.
The last term in Eq.\ (\ref{I}) is slightly different 
from the original functional of Eilenberger
where $g\hat{\bf v}$ appears in place of $(g\!-\!1)\hat{\bf v}$.\cite{Eilenberger68}
Although it does not change Eq.\ (\ref{Eilens}) at all,
it is found numerically that the modification is necessary for $F$ to have 
its absolute minimum with respect to $\Delta$, $\bf A$, and $f\!=\! f(\Delta,{\bf A})$
satisfying Eq.\ (\ref{Eilen}), 
as anticipated by Eilenberger.\cite{Eilenberger68}
It should be noted that Pesch and Kramer\cite{PK74} also 
adopted Eq.\ (\ref{F}) as a basis for their
numerical calculations.
More recently, Endres and Rainer\cite{ER03} performed a numerical calculation of the free energy
for both an SN contact and a single vortex based on
Eq.\ (\ref{F}), and compared the results with those from three free-energy functionals
obtained from the Luttinger-Ward functional.
They found numerical agreements among the values from the four different expressions.

Following Doria, Gubernatis, and Rainer,\cite{DGR89} I now rewrite
the right-hand side of Eq.\ (\ref{F}) in terms of 
\begin{eqnarray}
\begin{array}{c}
\vspace{2mm}
{\bf r}'\!\equiv{\bf r}/\lambda, \hspace{3mm}
B_{\lambda}\!\equiv\! \lambda^{2}B,\hspace{3mm}
\tilde{\bf A}_{\lambda}({\bf r}')\!\equiv \!\lambda\tilde{\bf A}(\lambda{\bf r}'),
\\
\Delta_{\lambda}({\bf r}')\!\equiv \!\Delta(\lambda{\bf r}'), \hspace{3mm}
f_{\lambda}(\varepsilon_{n},{\bf k}_{\rm F},{\bf r}')
\equiv f(\varepsilon_{n},{\bf k}_{\rm F},\lambda{\bf r}').
\end{array}
\label{DGRS}
\end{eqnarray}
I then differentiate the resulting expression with respect to $\lambda$
and put $\lambda\!=\! 1$.
Since the procedure (\ref{DGRS}) does not change the value of $F/V$,
we have $\frac{\partial}{\partial \lambda} (F/V)|_{\lambda=1}\!=\! 0$ from the left-hand side.
As for the right-hand side, the only implicit dependence to be considered is the one
from $B_{\lambda}$; those from $f_{\lambda}$,
$\Delta_{\lambda}$, and $\tilde{\bf A}_{\lambda}$ can be neglected 
due to the stationarity of Eq.\ (\ref{Eilens}).
We thereby obtain 
\begin{eqnarray}
&&\hspace{-5mm}0 = \left.\frac{\partial (F/V)}{\partial B_{\lambda}}
\frac{\partial B_{\lambda}}{\partial \lambda}\right|_{\lambda=1}-
2\kappa_{0}^{2}B^{2}-\frac{2\kappa_{0}^{2}}{V}\int\!d{\bf r}\,
({\bm \nabla}\!\times\!\tilde{\bf A})^{2}
\nonumber \\
&&-\frac{\pi T}{V}\!\sum_{n=0}^{\infty}\int \! d{\bf r} \left<\!
\frac{f^{\dagger}\hat{\bf v}\!\cdot\!({\bm \nabla}\!-\!i{\bf A})f
\!-\!f\hat{\bf v}\!\cdot\!({\bm \nabla}\!+\!i{\bf A})f^{\dagger}}{1+g}\!\right> .
\nonumber \\
\end{eqnarray}
Using the thermodynamic relation
$\frac{\partial (F/V)}{\partial B_{\lambda}}|_{\lambda=1}=\kappa_{0}^{2}H$
in the present units, we arrive at Eq.\ (\ref{H-B}).

\section{\label{sec:p-wave}Extension to the case with $p$-wave impurity scattering}

In the presence of $p$-wave impurity scattering, 
Eq.\ (\ref{Eilen}) is replaced by
\begin{eqnarray}
\left[\varepsilon_{n}+\frac{\langle g\rangle}{2\tau}
+\frac{d\,\hat{\bf k}\!\cdot\!\langle\hat{\bf k}'g\rangle}{2\tau_{1}}
+\frac{\hat{\bf v}}{2}\!\cdot\!\left({\bm \nabla}-i{\bf A}\right)\right] f
\nonumber \\=
\left(\Delta+\frac{\langle f\rangle}{2\tau}
+\frac{d\,\hat{\bf k}\!\cdot\!\langle\hat{\bf k}'f\rangle}{2\tau_{1}}\right)g \, ,
\label{Eilen-sp}
\end{eqnarray}
where $\langle\hat{\bf k}'g\rangle\!\equiv\!\langle\hat{\bf k}'
g(\varepsilon_{n},{\bf k}_{\rm F}',{\bf r})\rangle$, for example,
and $d\!=\!2,3$ is the dimension of the system.
This brings additional terms on the right-hand side of 
Eqs.\ (\ref{Eilen2-1}) and (\ref{Eilen2-3}) as
\begin{eqnarray}
\sum_{N'} \tilde{\cal M}_{NN'}\tilde{f}_{N'}^{(1)} &&\hspace{-3mm}= \delta_{N0}\left(1
+\frac{\langle \tilde{f}_{0}^{(1)}\rangle}{2\tau}
+\frac{d\cos\theta\, \langle \tilde{f}_{0}^{(1)}\!\cos\theta'\rangle}{2\tau_{1}}\right)
\nonumber \\
&&+\delta_{N1}\frac{d\sin\theta\,\langle \tilde{f}_{1}^{(1)}\!\sin\theta'\rangle}{4\tau_{1}}
\label{Eilen2-1p}
\end{eqnarray}
\begin{eqnarray}
&&\hspace{-3mm}\sum_{N'} \tilde{\cal M}_{NN'}\tilde{f}_{N'}^{(3)} 
=\delta_{N0}\left(\frac{\langle \tilde{f}_{0}^{(3)}\rangle}{2\tau}
+\frac{d\cos\theta\, \langle \tilde{f}_{0}^{(3)}\!\cos\theta'\rangle}{2\tau_{1}}\right)
\nonumber \\
&&\hspace{10mm}
+\delta_{N1}\frac{d\sin\theta\,\langle \tilde{f}_{1}^{(3)}\!\sin\theta'\rangle}{4\tau_{1}}
+J_{N}^{(3)}+J_{N}^{(A)} \, ,
\label{Eilen2-3p}
\end{eqnarray}
where $J_{N}^{(A)}$ is given by Eq.\ (\ref{J(A)}),
and $J_{N}^{(3)}$ is defined instead of Eq.\ (\ref{J(3)}) by
\begin{eqnarray}
&&J_{N}^{(3)}
\equiv-\frac{1}{2}\!\sum_{N'}(-1)^{N'} I_{NN'N+N'0}^{(4)}\,
\tilde{f}_{N'}^{(1)}\tilde{f}_{N+N'}^{(1)}
\nonumber \\
&&
\hspace{20mm}\times\biggl(\! 1
+\frac{\langle \tilde{f}_{0}^{(1)}\rangle}{2\tau}
+\frac{d\cos\theta\, \langle \tilde{f}_{0}^{(1)}\!\cos\theta'\rangle}{2\tau_{1}}\! 
\biggr)
\nonumber \\
&&+\frac{1}{4}\sum_{N'}(-1)^{N'}I_{NN'N'N}^{(4)}\,
\nonumber \\
&&
\hspace{10mm}\times\left(
\frac{\langle \tilde{f}_{N'}^{(1)}\tilde{f}_{N'}^{(1)}\rangle}{\tau}
+\frac{d\cos\theta\, \langle \tilde{f}_{N'}^{(1)}\tilde{f}_{N'}^{(1)}
\!\cos\theta'\rangle}{\tau_{1}}\right)\tilde{f}_{N}^{(1)}
\nonumber \\
&& 
+\frac{d\sin\theta}{8\tau_{1}}\sum_{N'}(-1)^{N'}\biggl[-I_{NN'N+N'-11}^{(4)}\,
\tilde{f}_{N'}^{(1)}\tilde{f}_{N+N'-1}^{(1)}
\nonumber \\
&&
\hspace{3mm}\times
\, \langle \tilde{f}_{1}^{(1)}\!\sin\theta'\rangle+
I_{NN'N'-1N+1}^{(4)}\,
 \langle \tilde{f}_{N'}^{(1)}\tilde{f}_{N'-1}^{(1)}
\!\sin\theta'\rangle\tilde{f}_{N+1}^{(1)}
\nonumber \\
&& \hspace{3mm}+I_{NN'N'+1N-1}^{(4)}\,
 \langle \tilde{f}_{N'}^{(1)}\tilde{f}_{N'+1}^{(1)}
\!\sin\theta'\rangle\tilde{f}_{N-1}^{(1)}\biggr]
 \, .
\label{J(3)p}
\end{eqnarray}

Equation (\ref{Eilen2-1p}) can be solved in the same way as Eq.\ 
(\ref{f(1)a}) to yield
\begin{eqnarray}
\tilde{f}_{N}^{(1)}={\tilde{K}^{0}_{N}}\left(1
+\frac{\langle \tilde{f}_{0}^{(1)}\rangle}{2\tau}
+\frac{d\cos\theta\, \langle \tilde{f}_{0}^{(1)}\!\cos\theta'\rangle}{2\tau_{1}}\right)
\nonumber \\
+
{\tilde{K}^{1}_{N}}\frac{d\sin\theta\, \langle \tilde{f}_{1}^{(1)}\!\sin\theta'\rangle}{4\tau_{1}}\, .
\label{f(1)p}
\end{eqnarray}
From Eq.\ (\ref{f(1)p}), we obtain self-consistent equations
for $\langle \tilde{f}_{0}^{(1)}\rangle$,
$\langle \tilde{f}_{0}^{(1)}\!\cos\theta\rangle$, and 
$\langle \tilde{f}_{1}^{(1)}\!\sin\theta\rangle$ as
\begin{eqnarray}
{\cal K}\left[\begin{array}{c}
\vspace{1mm}
\langle \tilde{f}_{0}^{(1)}\rangle\\
\vspace{1mm}
\langle \tilde{f}_{0}^{(1)}\!\cos\theta\rangle\\
\langle \tilde{f}_{1}^{(1)}\!\sin\theta\rangle
\end{array}
\right]
=
\left[\begin{array}{c}
\vspace{2mm}
\langle \tilde{K}_{0}^{0}\rangle\\
\vspace{2mm}
\langle \tilde{K}_{0}^{0}\!\cos\theta\rangle\\
\langle \tilde{K}_{1}^{0}\!\sin\theta\rangle
\end{array}
\right],
\label{<f(1)p>}
\end{eqnarray}
where the matrix ${\cal K}$ is defined by
\begin{equation}
{\cal K}\equiv\left[\!
\begin{array}{ccc}
\vspace{2mm}
\displaystyle
 1\!-\!\frac{\langle\tilde{K}_{0}^{0}\rangle}{2\tau}\!&\!
\displaystyle
 -\frac{d\langle\tilde{K}_{0}^{0}\!\cos\theta\rangle}{2\tau_{1}}\!&\!
 \displaystyle
 \frac{d\langle\tilde{K}_{1}^{0}\!\sin\theta\rangle}{4\tau_{1}}\\
\vspace{2mm}
\displaystyle
-\frac{\langle\tilde{K}_{0}^{0}\!\cos\theta\rangle}{2\tau}\!&\!
\displaystyle
1\!-\!\frac{d\langle\tilde{K}_{0}^{0}\!\cos^{2}\theta\rangle}{2\tau_{1}}\!&\!
\displaystyle
 \frac{d\langle\tilde{K}_{1}^{0}\!\sin2\theta\rangle}{8\tau_{1}}\\
\\
\displaystyle
-\frac{\langle\tilde{K}_{1}^{0}\!\sin\theta\rangle}{2\tau}\!&\!
\displaystyle
 -\frac{d\langle\tilde{K}_{1}^{0}\!\sin2\theta\rangle}{4\tau_{1}}\!&\!
 \displaystyle
1\!-\! \frac{d\langle\tilde{K}_{1}^{1}\!\sin^{2}\theta\rangle}{4\tau_{1}}
\end{array}
\!\right].
\label{calK}
\end{equation}
Noting $K_{N'}^{N}\!=\!K_{N'}^{N}(\tilde{\varepsilon}_{n},\beta)$ as seen from
Eqs.\ (\ref{R_N})-(\ref{KN'N2}) with $\beta$ defined in Eq.\ (\ref{beta}),
we immediately realize that 
$\langle \tilde{K}_{0}^{0}\!\cos\theta\rangle\!=\!0$ in Eq.\ (\ref{<f(1)p>}) 
and ${\cal K}_{2j}\!=\!{\cal K}_{j2}\!=\! 0$ for $j\!=\! 1,3$ in Eq.\ (\ref{calK}).
Thus, Eq.\ (\ref{<f(1)p>}) can be solved easily with 
$\langle \tilde{f}_{0}^{(1)}\!\cos\theta\rangle\!=\!0$.
Substituting the resulting expressions of 
$\langle \tilde{f}_{0}^{(1)}\rangle$ and 
$\langle \tilde{f}_{1}^{(1)}\!\sin\theta\rangle$ 
into Eq.\ (\ref{f(1)p}), we obtain
\begin{eqnarray}
&&\hspace{-4mm}
\tilde{f}_{N}^{(1)}
\nonumber \\
&&
\hspace{-4mm}=\frac{
\bigl(1
\!-\!\frac{d}{4\tau_{1}}{\langle \tilde{K}_{1}^{1}\!\sin^{2}\theta'\rangle}\bigr)\tilde{K}^{0}_{N}
+\frac{d}{4\tau_{1}}{\langle \tilde{K}_{1}^{0}\!\sin\theta'\rangle\tilde{K}^{1}_{N}\sin\theta}
}{
\bigl(1
\!-\!\frac{1}{2\tau}\langle \tilde{K}_{0}^{0}\rangle\bigr)\bigl(1
\!-\!\frac{d}{4\tau_{1}}\langle \tilde{K}_{1}^{1}\!\sin^{2}\theta'\rangle\bigr)
+\frac{d}{8\tau\tau_{1}}\langle \tilde{K}_{1}^{0}\!\sin\theta'\rangle^{2}}
\, .
\nonumber \\
\label{f(1)p2}
\end{eqnarray}
Equation (\ref{J(3)p}) is also simplified into
\begin{eqnarray}
&&\hspace{-3mm}J_{N}^{(3)}
\equiv-\frac{1}{2}\!\sum_{N'}(-1)^{N'} I_{NN'N+N'0}^{(4)}\,
\tilde{f}_{N'}^{(1)}\tilde{f}_{N+N'}^{(1)}
\biggl(\! 1
+\frac{\langle \tilde{f}_{0}^{(1)}\rangle}{2\tau}\! \biggr)
\nonumber \\
&&+\frac{1}{4\tau}\sum_{N'}(-1)^{N'}I_{NN'N'N}^{(4)}\,
\langle \tilde{f}_{N'}^{(1)}\tilde{f}_{N'}^{(1)}\rangle \tilde{f}_{N}^{(1)}
\nonumber \\
&& 
+\frac{d\sin\theta}{8\tau_{1}}\sum_{N'}(-1)^{N'}\biggl[-I_{NN'N+N'-11}^{(4)}\,
\tilde{f}_{N'}^{(1)}\tilde{f}_{N+N'-1}^{(1)}
\nonumber \\
&&
\hspace{3mm}\times
\, \langle \tilde{f}_{1}^{(1)}\!\sin\theta'\rangle+
I_{NN'N'-1N+1}^{(4)}\,
 \langle \tilde{f}_{N'}^{(1)}\tilde{f}_{N'-1}^{(1)}
\!\sin\theta'\rangle\tilde{f}_{N+1}^{(1)}
\nonumber \\
&& \hspace{3mm}+I_{NN'N'+1N-1}^{(4)}\,
 \langle \tilde{f}_{N'}^{(1)}\tilde{f}_{N'+1}^{(1)}
\!\sin\theta'\rangle\tilde{f}_{N-1}^{(1)}\biggr]
 \, .
\label{J(3)p2}
\end{eqnarray}

Equation (\ref{Eilen2-3p}) can be treated similarly to obtain
$\langle\tilde{f}_{0}^{(3)}\rangle$ which appears in Eq.\
(\ref{pair2}).
Then a straightforward calculation leads to the same expression
(\ref{f(3)}) for $\langle\tilde{f}_{0}^{(3)}\rangle$
with $\tilde{f}^{(1)}_{N}$ and $J^{(3)}_{N}$ replaced by Eqs.\ (\ref{f(1)p2})
and (\ref{J(3)p2}), respectively.
Another relevant quantity in Eq.\ (\ref{pair2}) 
is $\langle\tilde{f}_{0}^{(1)'}\rangle$, as seen from Eq.\ (\ref{f(1)expand}).
Differentiating Eq.\ (\ref{Eilen2-1p}) with respect to $B$ and 
solving the resulting equation self-consistently, we also obtain Eq.\ (\ref{f(1)'2})
for $\langle\tilde{f}_{0}^{(1)'}\rangle$
with $\tilde{f}^{(1)}_{N}$ in Eq.\ (\ref{J(2)}) replaced by Eqs.\ (\ref{f(1)p2}).

It hence follows that Eqs.\ (\ref{MainResults}) and (\ref{S}) remains the same
in the presence of the $p$-wave impurity scattering
with $\tilde{f}^{(1)}_{N}$ and $J^{(3)}_{N}$ replaced by Eqs.\ (\ref{f(1)p2}) and (\ref{J(3)p2}),
respectively.


\end{document}